\documentclass[journal]{IEEEtran}

\usepackage[pdftex]{graphicx}

\usepackage{array}
\usepackage{multirow}
\usepackage{bm}
\usepackage{mdwmath}
\usepackage{xcolor,soul,framed}
\usepackage{cite}
\usepackage{eqparbox}
\usepackage{url}
\usepackage{amsmath}
\usepackage{amssymb}
\usepackage{textcomp}
\usepackage{tikz}
\usepackage{bm}
\usepackage{dblfloatfix}

\newcommand{\branchesm}{\mathcal{E}}

\newcommand{\busesm}{\mathcal{N}}

\newcommand{\meas}{$\mathcal{M}$}
\newcommand{\measm}{\mathcal{M}}

\newcommand{\timeseries}{$\mathcal{T}$}
\newcommand{\timeseriesm}{\mathcal{T}}

\newcommand{\crmlijt}{\textbf{I}^{\text{re}}_{\text{lij,t}}}

\newcommand{\crmut}{\textbf{I}^{\text{re}}_{\text{u,t}}}
\newcommand{\cimlijt}{\textbf{I}^{\text{im}}_{\text{lij,t}}}

\newcommand{\cimut}{\textbf{I}^{\text{im}}_{\text{u,t}}}
\newcommand{\urjt}{\textbf{U}^{\text{re}}_{\text{j,t}}}
\newcommand{\urit}{\textbf{U}^{\text{re}}_{\text{i,t}}}
\newcommand{\uijt}{\textbf{U}^{\text{im}}_{\text{j,t}}}
\newcommand{\uiit}{\textbf{U}^{\text{im}}_{\text{i,t}}}

\newcommand{\p}{P}
\newcommand{\qm}{Q}

\begin{document}

\bstctlcite{IEEEexample:BSTcontrol}
   \title{Combined Unbalanced Distribution System State and Line Impedance Matrix Estimation}

\author{Marta~Vanin,~\IEEEmembership{Graduate Student Member,~IEEE, }
Frederik~Geth,~\IEEEmembership{Member,~IEEE,}
 Reinhilde~D'hulst, 
       and~Dirk~Van~Hertem,~\IEEEmembership{Senior Member,~IEEE}

\thanks{M. Vanin, and D. Van Hertem are with the Research Group ELECTA, Department of Electrical Engineering, KU Leuven, 3001 Heverlee, Belgium. 
R. D'hulst is with VITO, Boeretang 200, 3400 Mol, Belgium.
F. Geth is with GridQube, Springfield Central QLD 4300, Australia.
M. Vanin, D. Van Hertem and R. D'hulst are also with EnergyVille, Thor Park 8310, 3600 Genk, Belgium.
 Corresponding author: marta.vanin@kuleuven.be
 }
}
\maketitle

\begin{abstract}
To address the challenges that the decarbonization of the energy sector is bringing about, advanced distribution network management and operation strategies are being developed. Many of these strategies require accurate network models to work effectively. However, distribution network data are known to contain errors, and attention has been given to techniques that allow to derive improved network information. This paper presents a novel method to derive line impedance values from smart meter measurement time series, with realistic assumptions in terms of meter accuracy, resolution and penetration. The method is based on unbalanced state estimation and is cast as a non-convex quadratically constrained optimization problem. Both line lengths and impedance matrix models can be estimated based on an exact nonlinear formulation of the steady-state three-phase network physics. The method is evaluated on both the IEEE European Low Voltage feeder (906 buses) and a real Belgian LV feeder, showing promising results.  
\end{abstract}


\section{Introduction}\label{sec:introduction}

Distribution networks (DNs) are at the forefront of the energy transition, experiencing increased installations of rooftop PV, electric vehicles, and batteries. 
These bring both challenges and opportunities, and extensive research has been conducted on how to manage and integrate them in distribution system operations. 
However, it is well-known that historical distribution \emph{network models} are often inaccurate in terms of phase connectivity, and cable/overhead line types and lengths~\cite{nyserda}. Such errors in network data can compromise the reliability of simulation results on real-life DNs. 
Fortunately, smart meters (SMs) are increasingly available in DNs, to provide highly accurate measurements in terms of energy consumption for billing purposes. In many cases, they can also supply voltage, current, and/or power measurements, and may prove a powerful source of information to improve network models.

\subsection{Related Work}

This paper presents and compares methods to derive accurate impedance models for DNs, starting from SM measurement time series. 
Impedances, which are normally known inputs for power flow solvers, are \emph{unknowns} in this case, and estimation of impedance values is part of a class of problems known as the ``inverse power flow problem''\cite{9858017}. 
It has been established that this class of problems may not have a unique solution \cite{9858017,Li2019}, due to ``hidden nodes''. 

In the literature, impedance estimation (IE) is performed either stand-alone~\cite{ClaeysCIRED2021, Zhang2021}, or as part of broader-scoped topology identification (TI) workflows~\cite{Cunha2020}. Note that some TI methods might stop after the identification of the network's node and line connectivity, i.e., the topology itself, and do not necessarily include IE~\cite{Deka2020}. In other TI techniques, impedance values are assumed known and used as input~\cite{Cavraro2019}. A major difference between TI and IE is that the former is inherently a combinatorial problem, whereas the second is continuous. As such, the mathematical models and algorithms to address the two are usually fundamentally different, with different convergence and scalability properties. Theoretical and algorithmic considerations on the data-driven derivation of graph parameters are discussed in \cite{Li2019}, although these do not hold for the nonlinear SM measurements that are commonly available in European DNs.

Stand-alone IE examples aim to directly identify cable impedance values~\cite{Zhang2021}, or, alternatively, cable lengths~\cite{ClaeysCIRED2021}. The latter is less generic, as it assumes that the per length unit impedance of the lines are known. However, this assumption can be reasonable, especially in recently built systems or areas where overhead line or cable \emph{types} are strongly standardized.

To the best of the authors' knowledge, techniques in the literature on IE/TI are usually hardly applicable to generic real-world DNs, because of the following reasons:
\begin{enumerate}
    \item they consider balanced (positive-sequence) network models~\cite{Zhang2021, Short2013, Zhang2020Topology, Guo2022},
    \item they rely heavily on phasor measurement units (PMUs)~\cite{Li2022, GuptaPaolone}, or other advanced hardware capabilities like inverter probing~\cite{Cavraro2019CNS},
    \item they assume noiseless measurements~\cite{ClaeysCIRED2021, Guo2022}, and/or
    \item they assume a larger amount of measurement devices than realistically achievable, especially for LV DNs~\cite{Dutta}.
\end{enumerate} 
Combinations of the above are also present~\cite{Pegoraro2019, Lin2018}. Techniques for balanced networks might be appropriate for the transmission system, where the line parameters can also be inaccurate~\cite{Costa2022}, but DNs have nontrivial levels of phase unbalance. Methods have been devised specifically for North-American split-phase circuits~\cite{Lave2019, Short2013}. 

PMUs are accurate, high-resolution synchronized measurement devices that provide instantaneous voltage and current phasor measurements every few milliseconds. These allow to capture much more detailed information than averaged, low-resolution SM values. However, due to their cost, PMUs are rarely available in medium voltage (MV) DNs, and virtually absent on the LV side. 
Claeys et al.~\cite{ClaeysCIRED2021} show promising results for unbalanced PMU-free line length estimation. However,~\cite{ClaeysCIRED2021} uses assumption $3)$, which is unrealistic and facilitates the IE task with respect to a noisy scenario.

Works like~\cite{Dutta, WangPMAPS2020} present drawback $4)$. In~\cite{Dutta}, measurements are required at the both ends of every line. In~\cite{WangPMAPS2020}, a maximum-likelihood estimation method based on linearized power flow is presented. The IEEE 13-bus case is used to illustrate the method, and all electrically relevant buses seem monitored. 
Ref.~\cite{Li2022, Moffat2020} point out that measurement devices in DNs are usually only available at load nodes, which are usually the ``leaves" of the network tree. The methods in~\cite{Li2022, Moffat2020} show that leaf measurements can be enough, but they rely on $2)$, which is not economically feasible in DN due to the high cost of this equipment and the large scale of the system. When IE is performed with leaf SM measurements only, like in this paper, the problem is more degenerate, due to the existence of multiple solutions, and algorithmically more challenging.  

Several IE techniques are based on linear regression~\cite{Lave2019, Peppanen2016, Cunha2020}. Despite their simplicity, linear regression methods can provide high-quality results, but require considerable amounts of data~\cite{Peppanen2016}, i.e., long time series, and are subject to errors due to approximation of the actually nonlinear physics. 
Large amounts of data are also often a prerequisite of machine learning approaches: in~\cite{Yang2022} IE is solved with neural networks, with considerable training data requirements. 

Marulli et al.~\cite{Marulli2021} solve IE without $1)-4)$, but require measurements at each phase of MV/LV substations and, unlike state estimation (SE)-based techniques, do not filter the measurement noise, thus being more sensitive to measurement errors. Furthermore, examples like~\cite{Marulli2021, Cunha2020} calculate the positive sequence impedance only, instead of the full impedance matrix.
Techniques based on SE can and have been used for TI and IE before, e.g.~\cite{Jiang2020, bible, Eser2022, Karimi2021, Lin2018framework}. However, in a SE context, TI typically refers to the identification of the status of a limited amount of switches or portions of the network, and IE is generally not included~\cite{bible, Eser2022, Karimi2021}. Moreover, when IE is addressed with SE methods, it is only a few erroneous parameters that are estimated, in an otherwise well-known network~\cite{Lin2018framework}. This is not the case in DN, where the majority of the network data might be inaccurate. DN estimation is incorporated in a dynamic SE in~\cite{Jiang2020}. However, weather data are required, which might not be easily accessible. Finally, Shi et al.~\cite{Shi2011} calculate 3$\times$3 transmission line impedances with a SE-like approach that works well even for untransposed lines. However, the method would require PMUs at every line and both current and voltage phasor measurements and in this case the underlying SE reduces to a linear least square regression. As an alternative to SE-based techniques, Ban et al.~\cite{Jaepil2022} opt for reinforcement learning to mitigate the measurement noise. However, this requires a large measurement set to train the model, and the off-diagonal values of the impedance matrix are also not derived.

\subsection{Contributions and Assumptions}

We believe that there is a gap in the IE literature, as far as multi-conductor DNs without PMUs and limited SM availability are concerned, which we aim to address in this work. Therefore, two methods are presented to derive the line impedance values of DNs using only SM measurement time series, i.e., noisy voltage magnitude and active and reactive power measurements averaged over periods of fifteen minutes. The first method is dubbed ``line length estimation" (LLE), whereas the second, more generically, is ``impedance matrix estimation" (IME). Both build on a three-phase static SE framework, based on exact unbalanced power flow equations, which results in non-convex quadratic constrained programming (QCP) problems. As only leaf nodes are measured, multiple optimal solutions to the QCP may exist. In the IME case, constraints can optionally be added to enforce impedance matrix properties, such as diagonal dominance or symmetry. These constraints allow to reduce the search space of the QCP, reducing degeneracy and computational time, imposing realistic and/or desirable matrix properties. To the best of the authors' knowledge, this is the first paper that performs IME with impedance matrix structure.

The proposed methods assume that the phase connectivity of all users is known, as well as the general topology, i.e., the network tree. Reliable methods exist that perform phase~\cite{Hoogsteyn2022} and topology estimation~\cite{Cunha2020} with SM only, i.e., without relying on line impedance data. Furthermore, we develop a case study in which we recover the impedance of all lines from scratch, using historical measurements. Impedance values are assumed to be time-invariant. 

To summarize, this paper presents novel unbalanced SE-based IE approaches, whose advantages with respect to the state of the art are the following. Firstly, the measurement assumptions are exceedingly realistic, i.e., assumptions $1)-4)$ from the previous section are not required. Secondly, the exact steady-state power flow physics is leveraged, which provides accurate IE with relatively few time steps. Finally, the use of SE naturally reduces the impact of measurement errors on IE.

%
\section{Mathematical Programming Model}\label{sec:mathematical_model}
The proposed impedance identification method builds upon a SE model like the one presented in~\cite{Vanin2022}, and can be summarized as the generic optimization problem:
\begin{IEEEeqnarray}{ l C l }
    \text{minimize} \; \;       &~& \sum_{\substack{m \in \measm}} \sum_{\substack{t \in \timeseriesm}}   \rho_{m, t},         \label{eq:objective}    \\
    \text{subject to:}          &~& \mathbf{f}(\boldsymbol{\rho}, \mathbf{x}) = 0,   \label{eq:f}            \\
                                &~& \mathbf{h}(\mathbf{x}) = 0,                      \label{eq:h}            \\    
                                &~&  \mathbf{g}(\mathbf{x}) = 0,                     \label{eq:g}            \\
                                &~&  \mathbf{k}(\mathbf{x}) \leq 0. \label{eq:k} 
\end{IEEEeqnarray}
We can think of conventional SE as a single-period unbalanced OPF problem where we are minimizing a distance metric between observations and (functions of) state variables. For the combined SE and IE problem we are exploring, we 1) extend this to the multi-period case, and 2) make the impedance matrices variables (instead of them being knowns).

The objective \eqref{eq:objective} is to minimize the sum of a certain metric of all measurement residuals like in SE, but over a sequence of time steps. Let \meas \ be the set of all available measurements and \timeseries \ be that of the time steps with measurements. Each $m \in \measm$ is fully represented by $1)$ a real value $z_{m, t}$, i.e., the SM reading at time step $t$, $2)$ the system variable it refers to $x_{m,t}$, e.g., voltage magnitude of bus $i$, and $3)$ $\sigma_m$, which is a measure of the confidence on the measurement accuracy. In particular, as all measurement errors are assumed to follow a Gaussian distribution, $\sigma_m$ is one third of the maximum error for $m$, i.e., its standard deviation. This is a standard assumption in SE~\cite{Vanin2022}. The maximum error is provided on the SM specifications. In this work, SM are all assumed to belong to the 0.5\% accuracy class, which is in line with the capabilities of SM on the market~\cite{Vanin2022}, and is a more ``pessimistic" error assumption than for instance ~\cite{Cunha2020, Zhang2021}.

The residual formulation \eqref{eq:f} used in this paper is the exact linear relaxation of the weighted least absolute values~\eqref{eq:rWLAV1}-\eqref{eq:rWLAV2} (see \cite{Vanin2022}):
\begin{IEEEeqnarray}{rCl}
     \rho_{m, t} \geq \frac{x_{m,t} - z_{m,t}}{\sigma_m},\quad \forall m \in \measm, t \in \timeseriesm \label{eq:rWLAV1},\\
     \rho_{m, t} \geq - \frac{x_{m, t} - z_{m, t}}{\sigma_m},\quad  \forall m \in \measm, t \in \timeseriesm. \label{eq:rWLAV2}
\end{IEEEeqnarray}
Eq. \eqref{eq:h} represents Kirchhoff's current law (KCL) and multi-conductor Ohm's law. In this paper, we assume that the only available SM measurements are located at the users' premises, i.e., the ``leaf'' nodes of the tree. As such, all other nodes are zero-injection buses, which are fully captured as equality constraints by the KCL equations in \eqref{eq:h}. The current-voltage formulation is chosen to represent these physics, with a variable space consisting of branch currents and nodal voltages in rectangular coordinates.
The current-voltage formulation is exact for three-phase three-wire networks, i.e., it introduces no modelling error. This results in a numerically robust quadratically constrained programming problem, and has proved to solve efficiently with nonlinear programming methods implemented with common mathematical optimization toolboxes.

\begin{figure}[b]
    \centering
\includegraphics[trim={0 3.5cm 0 7.2cm},clip, width=5.6in]{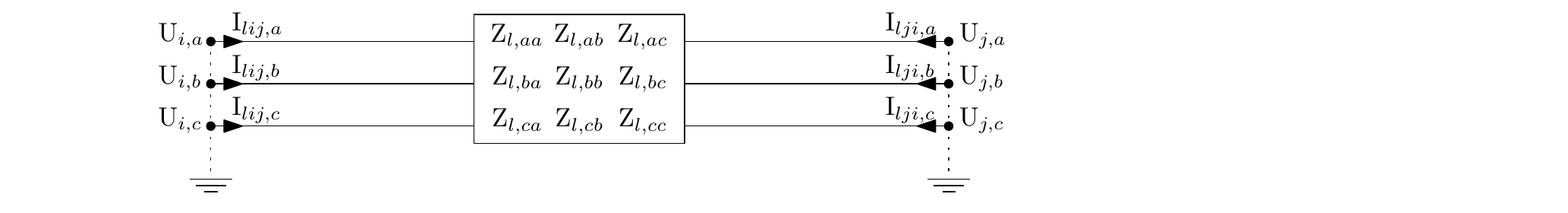}
    \caption[Branch model without shunt admittances]{Scalar variables for $3 \times 3$ branch model without shunt admittances, matching equations \eqref{eq:ohm_vcr}-\eqref{eq:ohm_vcr2}}
    \label{fig:impedance_noshunt}
\end{figure}

Let $\busesm$ be the set of the network buses, $\branchesm$ be that of its branches, $\mathcal{U}$ that of connected users (representing load/generation/storage/...). Every $l \in \branchesm$ connects two buses $i \in \busesm$ and $j \in \busesm$ and the \underline{\textbf{c}}onnectivity mapping is $lij \in \mathcal{C} \subseteq  \branchesm \times \busesm \times \busesm$. 
Every user $u \in \mathcal{U}$ is connected to a bus~$i \in \busesm$ and the connectivity mapping is $ui \in \mathcal{C}^{\text{u}} \subseteq  \mathcal{U} \times \busesm$.
In this paper, a bold typeface is used to indicate vectors and matrices. 
Vectors are used to stack phase values, i.e., for a generic  phasor quantity: $\mathbf{v} = [v_p]_{\forall p \in \Phi}$, where $\Phi \subseteq \{a,b,c\}$ are the phases that interest the quantity described by \textbf{v}.
Phasor $\crmut+\text{j} \cimut$ is the complex current from the bus towards user $u$ at bus $i$ at time $t$; $\crmlijt+\text{j} \cimlijt$ is the complex current in line $l$ in the direction of $i$ to $j$ at time $t$.
Note that the proposed notation works for meshed and radial feeders alike, and the index ``$l$" in ``$lij$" allows to differentiate between parallel branches. 

The multi-conductor KCL now reads as follows:
\begin{IEEEeqnarray}{l l}\label{eq:kcl}
    \sum_{lij \in \mathcal{C}} \! \crmlijt + \!\! \sum_{ui \in \mathcal{C}^{\text{u}}} \!\crmut = 0, &\quad
     \sum_{lij \in \mathcal{C}}\!
    \cimlijt + \sum_{ui \in \mathcal{C}^{\text{u}}} \! \cimut  = 0, \nonumber \\ 
   &\quad \forall t \in \timeseriesm,  i \in \busesm.
\end{IEEEeqnarray}
\noindent

The multi-conductor Ohm's law is:
\begin{IEEEeqnarray}{rCl}\label{eq:ohm_vcr}
    \urit = \urjt + \mathbf{R}_{\text{l}}  \crmlijt - \mathbf{X}_{\text{l}} \cimlijt,\; \forall lij \in \mathcal{C}, t  \in \timeseriesm, \\
    \uiit = \uijt + \mathbf{R}_{\text{l}} \cimlijt + \mathbf{X}_{\text{l}}  \crmlijt, \; \forall lij \in \mathcal{C}, t \in \timeseriesm, \label{eq:ohm_vcr2}
\end{IEEEeqnarray}
where $\urit+\text{j} \uiit$ is  bus voltage phasor at bus $i$ at time $t$, and $\mathbf{Z}_{\text{l}} = \mathbf{R}_{\text{l}} + \text{j} \mathbf{X}_{\text{l}} $ is the impedance matrix  of line $l$. 
Ohm's law is now a \emph{quadratic} equality, as $\mathbf{R}_{\text{l}}, \mathbf{X}_{\text{l}}$ are (linear transformations of) decision variables in the proposed optimization problem. This is the key generalization with respect to canonical unbalanced SE and OPF formulations like those presented in~\cite{GethIVR}. Note that the voltage and current phasors are time-dependent (with corresponding index $t$), whereas the impedance matrix variables only exist once for each line.

In the network model used in this work, impedances  $\mathbf{Z}_{\text{l}}$ are either scalars (for single-phase user lines), or matrices of dimension $3 \times 3$, as illustrated in Fig.~\ref{fig:impedance_noshunt}. This allows to capture many real-world network types. We neglect the shunt admittance of the lines/cables, a common assumption that has been shown to introduce negligible errors in LVDNs~\cite{Urquhart2015}.
Eq.~\eqref{eq:g} represent ``mapping" functions that are required to incorporate measurements which do not refer to quantities in the formulation's variable space. In this case, for every \emph{measured} user the following apply:
\begin{eqnarray}\label{eq:square}
    &  (\mathbf{U}^{\text{mag}}_\text{i,t})^2  = (\uiit)^2 + (\urit)^2 \;\; \forall i  \in \busesm, \forall t \in \timeseriesm, \label{eq:vm_meas}\\
     & \mathbf{\p}_\text{u,t} = \urit \circ \crmut + \uiit \circ \cimut \;\; \forall ui \in \mathcal{C}^{\text{u}}, \forall t \in \timeseriesm, \label{eq:p_meas}\\
     & \mathbf{\qm}_\text{u,t} = \uiit \circ \crmut - \urit \circ \cimut \;\; \forall ui \in \mathcal{C}^{\text{u}}, \forall t \in \timeseriesm, \label{eq:q_meas}
\end{eqnarray}
where $ \mathbf{\p}_\text{u,t} ,\mathbf{\qm}_\text{u,t}$ are the measured active and reactive power demand or generation, $\mathbf{U}^{\text{mag}}_\text{i,t}$ the measured voltage magnitude, and $\circ$ indicates element-wise multiplication.
Eq.~\eqref{eq:k} indicate the generic set of all variable bounds and (in)equality constraints that need to be added to complete the problem. In this specific case, these have been used to impose structural properties to the impedance matrices in the IME part of this work, as reported in Section~\ref{sec:ime}. Finally, note that the result of \eqref{eq:objective}-\eqref{eq:k} provides both the (time-invariant) impedance and the state variables for all time steps.

\subsection{Line Length Estimation (LLE)}\label{sec:lle}

In regular (optimal) power flow or SE calculations, $\textbf{Z}_\text{l}$ in \eqref{eq:ohm_vcr} is given as input, whereas here it is a variable which is solved for simultaneously with the multi-period SE. In the case of LLE, it is assumed that the per unit length cable parameters  $\mathbf{R}_{\text{l}}^{\text{nom}}, \mathbf{X}_{\text{l}}^{\text{nom}} $ are known, and only the line length $\ell_{l}$ is unknown, which is represented as:
\begin{eqnarray}\label{eq:impedance_scaled_length}
\mathbf{R}_\text{l} = \ell_{l} \cdot \mathbf{R}^\text{nom}_\text{l}, \quad
\mathbf{X}_\text{l} = \ell_{l} \cdot \mathbf{X}^\text{nom}_\text{l}, \quad\forall  l \in \branchesm.
\end{eqnarray}
Furthermore, if it can be assumed that the modeller knows that a line length $\ell_{l}$ is in a certain range $\underline{\ell_{l}} \leq \ell_{l} \leq \overline{\ell_{l}}$, this can be added as likelihood information in \eqref{eq:objective} as:
\begin{equation}
    \rho_{l, t} \geq \frac{(\ell_{l} - \widehat{\ell_{l}}) \cdot 3}{\overline{\ell_{l}}-\underline{\ell_{l}}}, \; \; \forall t \in \timeseriesm, l \in \branchesm, \label{eq:llikelihood1}
\end{equation}
\begin{equation}
    \rho_{l, t} \geq -\frac{(\ell_{l} - \widehat{\ell_{l}}) \cdot 3}{\overline{\ell_{l}}-\underline{\ell_{l}}}, \; \; \forall t \in \timeseriesm, l \in \branchesm, \label{eq:llikelihood2}
\end{equation}
where $\widehat{\ell_{l}}$ is a guess on the line length, within the given range.
Lengths are time-invariant, but residuals are not. The factor ``$\cdot 3$" in \eqref{eq:llikelihood1}-\eqref{eq:llikelihood2} stands for the third of the maximum length error, i.e., the standard deviation of the error distribution, assuming this is Gaussian. Even though it is often an approximation, Gaussian error distributions are the standard assumption in SE~\cite{bible}. 
Note that this is a similar but more generic version of the LLE method proposed by Claeys et al.~\cite{ClaeysCIRED2021}.

\subsection{Impedance Matrix Estimation (IME)}\label{sec:ime}

For IME, the optimization problem is  \eqref{eq:rWLAV1}-\eqref{eq:square}, where $\mathbf{R}_\text{l}, \mathbf{X}_\text{l}$ in \eqref{eq:ohm_vcr} are matrices composed of variables. In this case, matrix entries are \emph{not} added to the residuals, as it is impractical to guess or limit the value of each entry. The constraints below are added, to impose some structural properties to the impedance matrices. 

\subsubsection*{Non-negativity}

\begin{equation}\label{eq:positivity}
    R_{l, pq} \geq 0, X_{l, pq} \geq 0,  \; \; \forall p,q \in \{a, b, c \}, l \in \branchesm.
\end{equation}

Note that the physical properties of the lines imply strict positivity, but we allow zero values for simplicity. 

\subsubsection*{Symmetry constraint}

\begin{equation}\label{eq:symmetry}
    Z_{l, pq} = Z_{l, qp}, \; \; \forall p,q \in \{a, b, c \}, l \in \branchesm.
\end{equation}

\subsubsection*{X/R ratio}

\begin{equation}\label{eq:xrratio}
    \alpha_{l,2} \cdot X_{l, pq} \leq R_{l, pq} \leq \alpha_{l,1} \cdot X_{l, pq}, \; \; \forall p,q \in \{a, b, c \}, l \in \branchesm,
\end{equation}

where $\alpha_{l,1} \geq \alpha_{l,2} \geq 1$.

\subsubsection*{Diagonal dominance constraint}

\begin{equation}\label{eq:diagonal_dom}
    |Z_{l, pp}| \geq \sum_{p \neq q} |Z_{l, pq}|,\; \; \forall p,q \in \{ a, b, c\}, l \in \branchesm.
\end{equation}

Row diagonal dominance is enforced in this case, but since the impedance matrices are symmetrical, column diagonal dominance would be equivalent. Also, it can be applied to $\mathbf{Z}_{\text{l}}$ as above, or to $\mathbf{R}_{\text{l}}$ or $\mathbf{X}_{\text{l}}$ only.

\subsubsection*{Off-diagonal ratio}

\begin{equation}\label{eq:offdiagonal_ratio}
 \alpha_{l,4} \cdot Z_{l, pq} \leq Z_{l, pp} \leq \alpha_{l,3} \cdot Z_{l, pq}, \; \; \forall p,q \in \{a,b,c\}, l \in \branchesm.
\end{equation}
where $\alpha_{l, 3} \geq \alpha_{l, 4} \geq 1$.

\subsubsection*{Equal diagonal constraint}

\begin{equation}\label{eq:equal_diagonal}
    Z_{l, aa} = Z_{l, bb} = Z_{l, cc}, \forall l \in \branchesm.
\end{equation}

\subsubsection*{Equal off-diagonal}

\begin{equation}\label{eq:equal_offdiagonal}
    Z_{l, ab} = Z_{l, ac} = Z_{l, bc}, \forall l \in \branchesm.     
\end{equation}

\begin{figure}[b]
    \centering
\includegraphics[width=0.5\linewidth]{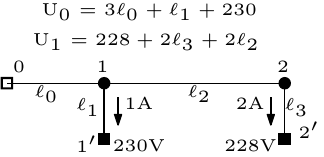}
    \caption[LLE with noiseless measurements]{LLE with noiseless measurements on a DC circuit: $\textrm{U}_{1^\prime}$~=~230\,V, $\textrm{U}_{2^\prime}$~=~228\,V, $\textrm{I}_{1^\prime}$ = 1\,A, $\textrm{I}_{2^\prime}$ = 2\,A. Example of multiple solutions: $\ell_{0}~=~1, \ell_{1}~=~1$ and $\ell_{0} = 1/3, \ell_{1}~=~3$.}
    \label{fig:multiple_solutions}
\end{figure}

\begin{figure*}[t]
    \centering
\includegraphics[width=1\textwidth]{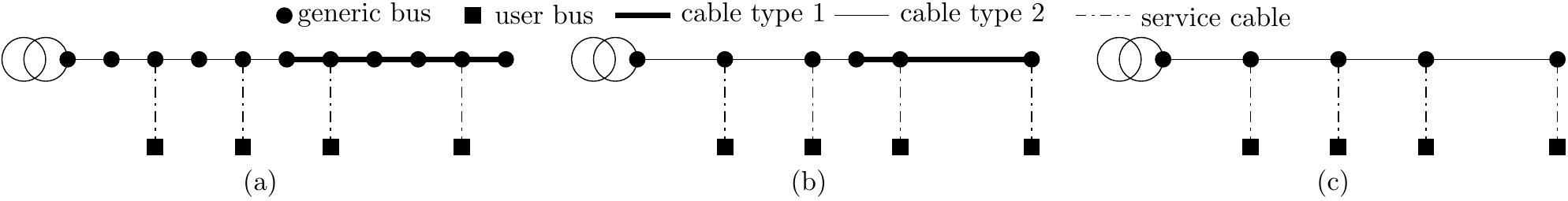}
    \caption{Original (a) and reduced (b)-(c) DN data. Reduction (b) preserves different cable type information, while (c) does not.}
    \label{fig:bus-reduction}
\end{figure*}

Note that constraint \eqref{eq:equal_offdiagonal} is only defined for off-diagonal elements of the upper diagonal part of the matrix. This is because, combined with the symmetry constraint \eqref{eq:symmetry}, this applies to the lower diagonal as well. Explicitly defining the same for the lower triangular part of the matrix would lead to redundant constraints, which may deteriorate the algorithmic performance. In this paper, the constraints above are applied to all branches, but nothing prevents to limit them to a subset of the branches if appropriate. Furthermore, all the constraints above are linear.  Matrices $\mathbf{R}_{\text{l}}$ and $\mathbf{X}_{\text{l}}$ for multiconductor lines are expected to be both symmetric and positive-definite (non-superconducting) due to physics (all lines are lossy) and as confirmed in Carson's and Polaczek's equations to obtain impedance values from geometry and material characteristics. Symmetry is easy to represent in optimization problems, while positive-definiteness is not. Semi-definite programming solvers, e.g., MOSEK, do not accept non-convex constraints, and nonlinear programming solvers, e.g., \textsc{Ipopt}, do not have positive-semi-definite constraint primitives. While not fundamental for the present work, it is interesting to note that the combination of positive diagonal elements \eqref{eq:positivity}, diagonal dominance \eqref{eq:diagonal_dom} and symmetry \eqref{eq:symmetry} is equal to linearly enforcing that the matrix is positive semi-definite.


\subsection{Multiple Solutions, Data Preprocessing and Validation}

In both the LLE and IME approaches, we end up with feasible sets that are non-convex quadratic. To obtain numerical results, we hand over the model to a derivative-based nonlinear optimization solver, and choose the open-source \textsc{Ipopt}~\cite{ipopt}. 
It is noted that the nonlinear programming routines implemented in \textsc{Ipopt}, when applied to non-convex problems, only have local convergence and optimality. 
Despite those relatively weak theoretical guarantees, such routines have proven very reliable in the context of OPF \cite{kardos2018complete}, when care is taken to avoid implementation pitfalls, e.g. establishing valid lower bounds on voltage magnitude  throughout the network \cite{geth2022pitfalls}. 
Global optimization methods applied to large OPF case studies \cite{gopinath2022benchmarking} have generally not been able to show that \textsc{Ipopt} and other solvers converge to truly local solutions. 
That being said, the authors are aware that these positive experiences may not generalize to the problems formulated in this paper. 
The only equation impacted by the OPF generalization proposed here is Ohm's law \eqref{eq:ohm_vcr}-\eqref{eq:ohm_vcr2}, due to it becoming quadratic.
In the authors' experience, it is best to verify this doesn't lead to degeneracy or  a disconnected search space, due to the bilinear term including complementary solutions ($ \mathbf{R}_{\text{l}} +\text{j}  \mathbf{X}_{\text{l}}$ and $\crmlijt+\text{j} \cimlijt$ both have zero as a feasible solution). 
Therefore, we tested the behavior using valid strictly positive lower bounds on $R_{l, pq}, X_{l, pq}$ instead of nonnegativity \eqref{eq:positivity}, however we failed to observe any significant changes in convergence or optimality. 
These observations on the quality of the solution need to be complemented to that of its uniqueness, as done hereafter.

Fig.~\ref{fig:multiple_solutions} shows the existence of multiple solutions in a very simplified LLE case: a purely resistive single-phase five-bus network, assuming for simplicity that there are no line losses, that R is unitary and that voltage and current measurements for users at buses $1^\prime$ and $2^\prime$ are available. Even assuming noiseless measurements, multiple line length values combinations exist, that return the same state $U_0, U_1$. Furthermore, DN data are often derived from paths in a geographical information system, resulting in the presence of large numbers of electrically superfluous nodes, as shown in Fig.~\ref{fig:bus-reduction} (a). To reduce the degeneracy, it is necessary to remove the superfluous nodes. This can be done as shown in Fig.~\ref{fig:bus-reduction} (b)-(c) and does not affect the result accuracy. As we are not interested in preserving cable type information, (c) is used.

In the LLE problem, the impedance is parameterized by one scalar variable per branch $\ell_{l}$, and it is reasonably easy to bound it within a certain length range. For the IME problem, we explore the ``transposed", ``untransposed" and ``diagonal" cases. 
In distribution networks, lines and cables are rarely transposed in the real world due to cost implications, i.e., additional effort, so the untransposed model is preferable. 

The authors note that positive and zero sequence parameterizations are commonly used by DSOs to model LV grid impedances~\cite{TaxonomyStudy}, which leads to perfectly transposed line/cable models\footnote{Furthermore, in 3P+N networks, one can only physically achieve such balanced network impedances through transposition across the length of the conductor, as it is impossible to obtain a symmetric geometry for four conductors.}. This is an approximation, and implies that 1) the corresponding impedance matrix in phase coordinates is balanced (same value for all diagonal entries, and same values for all off-diagonals), and 2) has been through the process of Kron's reduction of the neutral under the assumption that the neutral voltage is (close to) 0\,V. 
While we do not believe that representing three-phase four-wire lines with $3\times 3$ impedance models is always appropriate\footnote{(although reduced impedance models for sparsely grounded four-wire networks do exist~\cite{GethACM})}, we take this assumption in this work to reduce the amount of variables necessary for the impedance matrix parameterization. Furthermore, if reduced models are preferred, e.g.,  for computational reasons, the proposed method can also be used to ``learn" good $3 \times 3$ impedance models, in conditions where the Kron or other reductions are inexact. This is because 1) the proposed method can be seen as some form of nonlinear physics-informed regression, and 2) given the lack of a unique solution there is some ``slack'' to distribute impedance values in a way that they become a better ``fit".


If impedance matrices are for transposed lines, the diagonal values are equal, and off-diagonal values are equal, so constraints \eqref{eq:equal_diagonal}-\eqref{eq:equal_offdiagonal} match reality. As such, there are ``only" four scalar variables for each branch (resistance/reactance $\times$ diagonal/off-diagonal). 
In the untransposed case, symmetry is preserved, but diagonal and off-diagonal values can differ, leading to twelve scalar variables for each $3 \times 3$ matrix. 
This, combined with the fact that there are multiple possible solutions makes the IME problem more prone to algorithmic convergence issues, especially in the untransposed case.
Finally, the diagonal case enforces zero mutual impedance values, but does allow for different diagonal entries, so that there are a total of six scalar variables for each $3 \times 3$ matrix. This is an approximation, as mutual impedances are not zero, and is used to mimic the results of existing single-phase impedance estimation methods, e.g.,~\cite{Peppanen2016}, when applied to each phase circuit separately (note that, however, with respect to~\cite{Peppanen2016, Lave2019} and similar, we preserve non-convexity instead of linearizing the physics).

The existence of multiple solutions for the same fitting criterion makes it impossible to compare the original feeder impedances/lengths to the estimated ones, and two other metrics are used to assess the IE performance. The first is the estimated \textit{cumulative impedance} accuracy of every user, as proposed in~\cite{Marulli2021}. This consists of the error on the \textit{sum of the self-impedances of each user's path to the root node}\footnote{E.g., the cumulative resistance for a user at node 2$^\prime$ in Fig.~\ref{fig:multiple_solutions} is $R_0+R_2+R_{2^\prime}$, for a user at 1$^\prime$, this is $R_0+R_{1^\prime}$.}. 
The second metric is the power flow (PF) validation, as in~\cite{ClaeysCIRED2021}. In this case, we split the measurement time series into training and validation sets, and calculate the difference in PF results on the validation set using the original network data and the estimated one based on the training set. 


%
\section{Case Study Data and Setup}\label{sec:case_studies}
\begin{table*}[t]
\centering
\caption{Feasible sets for the three optimization problem variants: LLE, IME transposed, and IME untransposed} \label{tab:feasiblesets}
\begin{tabular}{l c c c c} 
\hline
 & LLE & IME transposed & IME untransposed & IME diagonal \\ 
\hline
Objective & \eqref{eq:objective} 
& \eqref{eq:objective} &\eqref{eq:objective}  &\eqref{eq:objective}  \\\hline
Variables & $ \rho_{m, t}, \urit, \uiit, \crmlijt, \cimlijt $  & $ \rho_{m, t}, \urit, \uiit ,  \crmlijt, \cimlijt$  & $ \rho_{m, t}, \urit, \uiit,  \crmlijt, \cimlijt $ & $ \rho_{m, t}, \urit, \uiit,  \crmlijt, \cimlijt $ \\
     &$\underline{\ell_{l}} \leq \ell_{l} \leq \overline{\ell_{l}}$  & $Z_{l, aa}, Z_{l, ab}$  & $Z_{l, aa}, Z_{l, ab}, Z_{l, ac}, Z_{l, bb}, Z_{l, bc}, Z_{l, cc}$ & $Z_{l, aa}, Z_{l, bb}, Z_{l, cc} $ \\ \hline
KCL & \eqref{eq:kcl} &  \eqref{eq:kcl} &  \eqref{eq:kcl} &  \eqref{eq:kcl} \\
Ohm's law & \eqref{eq:ohm_vcr}-\eqref{eq:ohm_vcr2}  & \eqref{eq:ohm_vcr}-\eqref{eq:ohm_vcr2} & \eqref{eq:ohm_vcr}-\eqref{eq:ohm_vcr2} & \eqref{eq:ohm_vcr}-\eqref{eq:ohm_vcr2} \\
Impedance structure & \eqref{eq:impedance_scaled_length} & \eqref{eq:positivity}-\eqref{eq:equal_offdiagonal} & \eqref{eq:positivity}-\eqref{eq:offdiagonal_ratio} & \eqref{eq:positivity}, \eqref{eq:xrratio}, $Z_{l,pq} = 0$  \\
Auxiliary residuals & \eqref{eq:llikelihood1}-\eqref{eq:llikelihood2} & - & - & -\\
Measurement functions & \eqref{eq:vm_meas}-\eqref{eq:q_meas} & \eqref{eq:vm_meas}-\eqref{eq:q_meas} & \eqref{eq:vm_meas}-\eqref{eq:q_meas} & \eqref{eq:vm_meas}-\eqref{eq:q_meas} \\
 \hline
\end{tabular}
\end{table*}

In this section, four impedance estimation problems are tested on the IEEE European LV Test Feeder (ELTF), as it is probably the best-known feeder of the ENWL DN library~\cite{ENWLreport}, and on a real LV feeder. Table~\ref{tab:feasiblesets} summarizes the feasible sets of the four impedance estimation problem variants: LLE, IME with transposed network data, IME with untransposed network data and the IME diagonal case.

\begin{figure*}[t]
    \centering
\includegraphics[width=\linewidth, trim={0 0 0 7.2cm},clip]{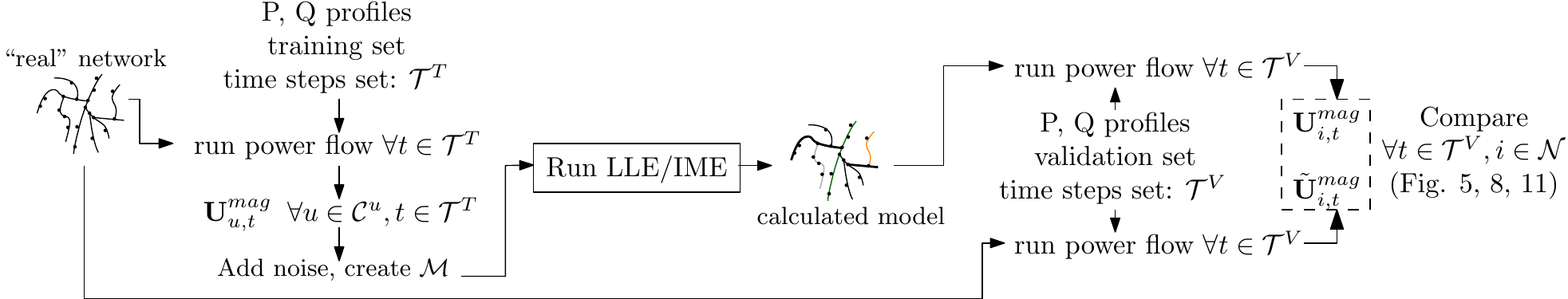}
    \caption[Flow of the LLE/IME calculation and validation process]{Flow of the LLE/IME calculation and validation process. Note that the ``real" network is transposed for LLE and transposed IME, and untransposed for untransposed IME calculations.}
    \label{fig:ime_lle_flow}
\end{figure*}

\subsection{Properties of cable types in the ENWL DN library}
This library features 128 3P+N LV feeders and five minute active power profiles for LV users, including both ``basic" demand and patterns for low-carbon technologies. Impedance values of the DN are calculated from 201 unique cable types (positive and zero sequence impedance pairs), which are also part of the data set.
Note that this data set, in its original version, therefore features transposed line models only. As such, constraints \eqref{eq:symmetry}, \eqref{eq:equal_diagonal} and \eqref{eq:equal_offdiagonal} always hold. Furthermore, $\mathbf{X}_\text{l}$ are diagonally dominant for more than 99\% of the whole data set cable length, and $\mathbf{R}_\text{l}$ for $\approx$96\% thereof. Finally, the following $\alpha_{l}$ bound values have been chosen for the \textit{\textbf{transposed}} IME, as they are never exceeded in the \textit{\textbf{transposed}} data set:
\begin{itemize}
    \item $\alpha_{l,1}$: 16 for service cables, 35 and 130 for the diagonal and off-diagonal elements of three-phase cables. 
    \item $\alpha_{l,2}$: 8.96 for service cables, otherwise 1.1 and 2.
    \item $\alpha_{l,3}$: 14 for X, 2 for R.
    \item $\alpha_{l,4}$: 50 for X, 70 for R.
\end{itemize}
Note that the cable types used in the ELTF are a subset of those of the ENWL dataset, and we assume that we do not know which ones are there in the feeder. It is often the case that the cable types present in a certain area are limited, and there are a finite number of cable manufacturers and types. As such, similar a-priori analyses should typically be possible, at least for transposed cases. In our \textit{\textbf{untransposed}} case, we do not use these bounds, as they are less practical to calculate and apply when all the matrix entries are different (up to symmetry). 
In the \textit{\textbf{untransposed}} case we replace the original ELTF impedance parameters with values from the open finite-element-derived cable dataset by Urquhart et al. \cite{Urquhart2015} using the same mapping methodology presented in Claeys et al. \cite{CLAEYS2022108522}. 
We only add bounds on the R/X and offdiagonal ratios for the case with 200 time steps, as this otherwise does not converge within the cut-off time of 24 hours. These bounds are the same for every applicable matrix entry, and are much ``looser" than the $\alpha$'s above ($\approx 100/0.01 \times$ the highest/lowest ratios found in the untransposed dataset), but was enough to make the untransposed IME converge. 

For the real LV feeder used in this paper, we do not make a-priori assumptions on the cable types for the IME analysis.

Note that in the real world often only \emph{part} of the line types are unknown, whereas in both our test cases we assume that all of them are. If some impedances are known, the respective variables can be fixed, and estimating the remainder parts only would be easier and faster, as extra system information is available and the problem is smaller. The presented results are thus essentially a worst-case scenario with respect to assumptions on the network data knowledge.

\subsection{European LV Test Feeder (ELTF) and Power Profiles}
The ELTF has 55 single-phase users and is unbalanced. The ``main" cable is modelled as three-phase, whereas all service cables to the users are modelled as single-phase cables. Note that only the LV network is modelled, i.e., the MV/LV transformer model is not included, nor is a MV network. With respect to other types of calculations, e.g., power flow analysis, no approximation is introduced by decoupling the MV and LV: the LV part is fully observable stand-alone, from a DSSE perspective, and so would be a MV network with the same measurement assumptions. Decoupling the two allows to perform IE in parallel on the different LV or MV networks. To explicitly address MV+LV systems, transformer equality constraints can be added, which is straightforward to do with the underlying DSSE framework~(see \cite{Vanin2022}). However, utilities typically have wrong MV/LV models, e.g., wrong tap settings, nominal impedance, etc., so explicitly including transformer constraints might actually introduce extra errors, leading to overall worse results unless identification of transformer parameters is performed beforehand.
 
 Before pre-processing, the ELTF has 906 buses. After the pre-processing step in Fig.~\ref{fig:bus-reduction} (c), the buses are reduced to 109 and the branches to 108. It is assumed that it is winter and all users have a heat pump, and the corresponding power profiles are extracted from the ENWL data set. This scenario presents limited voltage drops, which limits both the usefulness and the performance of IME. Therefore, user loads are scaled by a factor of three. This provides more ``information", as small power flows would otherwise cause small voltage drops that are indistinguishable from measurement noise. Scaling was not required for LLE: providing the nominal matrix values compensates for the lack of information due to the the lack of voltage drops.

As previously shown, we neglect cable shunt admittances, which leads to negligible errors for LVDN like the one used in this test case. For MVDNs, shunt currents might not be negligible in low-load conditions~\cite{kersting_distribution_2002}. However, at low load, the network is unlikely to show problematic behaviour or significant voltage sags, resulting in limited information as the network essentially behaves like a copper plate. Thus, as we are most interested in high load conditions, neglecting shunt admittances may prove acceptable for MVDNs too. 

The case study set-up and PF validation are illustrated in Fig.~\ref{fig:ime_lle_flow}. Each SM can measure P, Q and $U^{\text{mag}}$, which is a reasonable real-world set-up. To generate measurements from the ENWL active power data, $\textrm{Q }=\textrm{P}\cdot \tan\phi$ is assumed, with $\cos\phi~=~0.97$, and PF calculations are run. Gaussian errors are added on P and Q and the $U^{\text{mag}}$ resulting from the PF. Then, these five-minute noisy measurements are aggregated into fifteen minute averaged values. Data for 200 fifteen-minute time steps is generated, from which the N time steps with largest voltage drops have been selected, as reported in the x-axis of the result plots, to show how the estimation improves, in general, if more time steps are provided. Finally, ten extra time steps are used as validation set for PF calculations. The time steps, in this exercise, can be seen as independent, non-consecutive samples from the historical measurement dataset.

 \begin{figure}[b]
\includegraphics[width=0.45\textwidth]{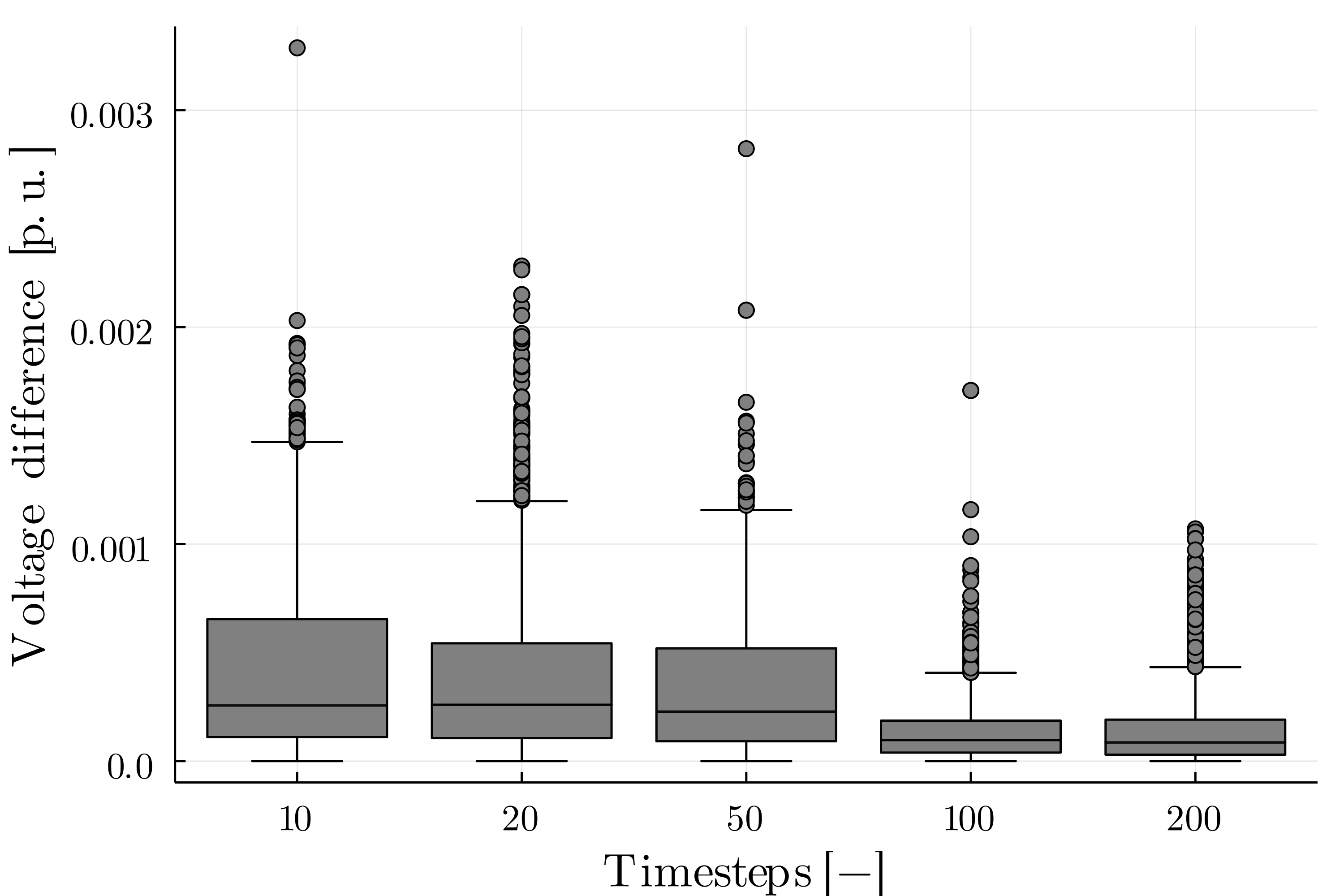}
\caption{Difference between the PF result with the original and estimated lengths (LLE case). Each box plot has 89\,100 entries: 55 single-phase buses $\times$ 54 three-phase buses $\times$ 10 PFs.}
\label{fig:lle_pf}
\end{figure}

\begin{figure*}[t!]
\begin{tabular}{cc}
  \includegraphics[width=0.45\textwidth]{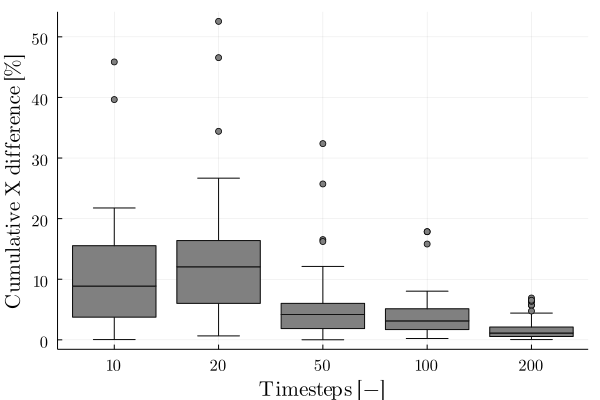}\label{fig:lle_X} &
  \includegraphics[width=0.45\textwidth]{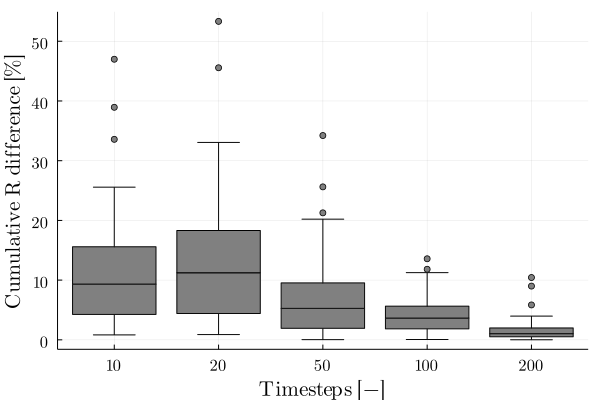}\label{fig:lle_R} \\
\end{tabular}
\caption{Difference between the users' actual and LLE-calculated cumulative reactance (left) and resistance (right). The boxplots have one entry for each user. Plot for $|$Z$|$ is omitted as it is similar to that of R (because X only contributes to a small part of $|$Z$|$). }
\label{fig:lle_xr}
\end{figure*}

 \begin{figure}[t]
 \centering
\includegraphics[width=0.40\textwidth]{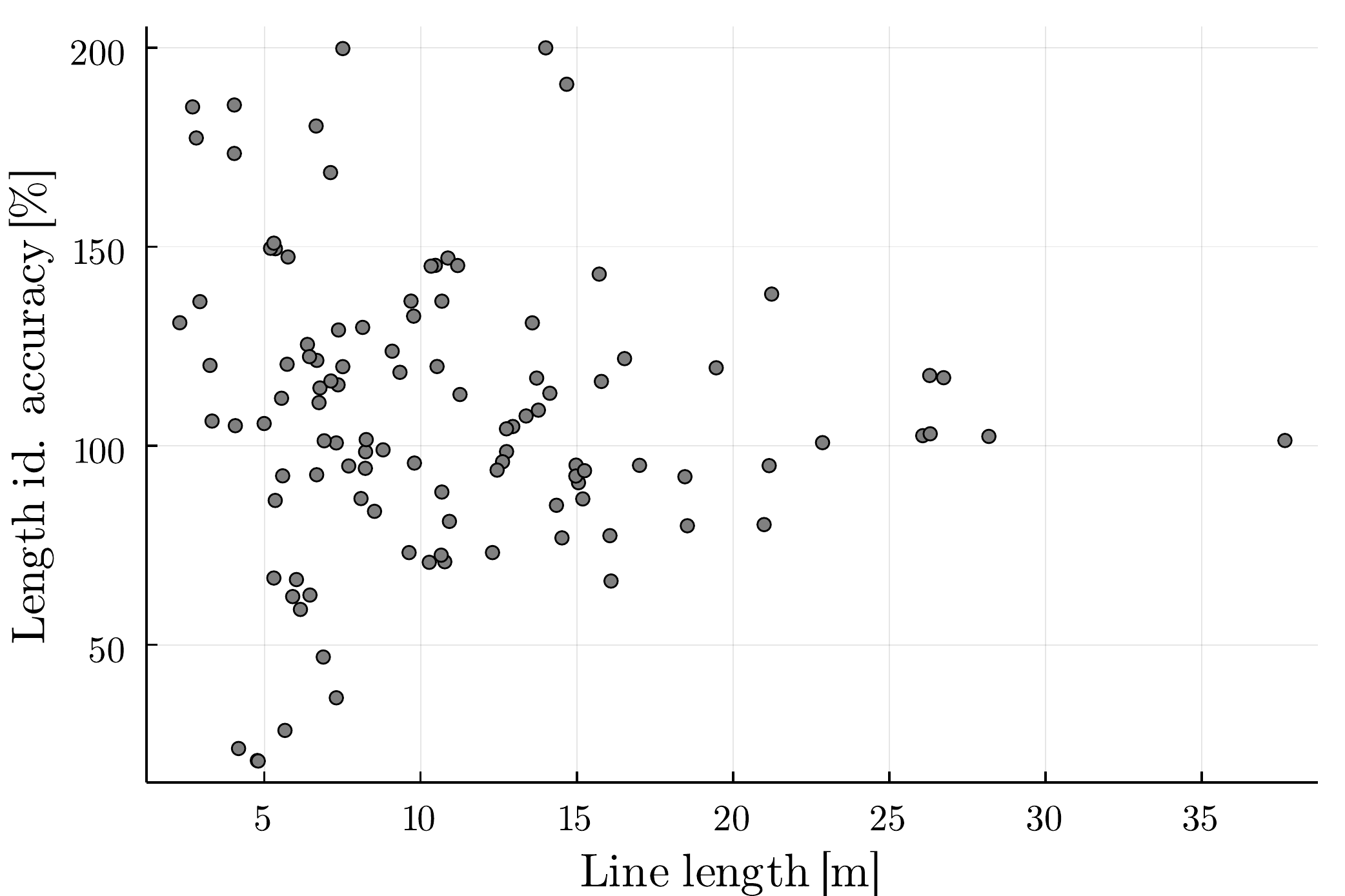}
\caption{Estimation accuracy of the individual line lengths, for 200 time steps.}
\label{fig:wrong_lengths}
\end{figure}

The problem is implemented with \textsc{JuMP}~\cite{JuMP} and solved with \textsc{Ipopt} v3.12.10~\cite{ipopt}, using MUMPS as underlying linear solver and tolerance 1E-7. Simulations are run on an Intel Xeon CPU E5-4610 v4, 32 GB RAM, using \textsc{Julia} 1.6, \textsc{PowerModelsDistributionStateEstimation} 0.6.0~\cite{Vanin2022}, and \textsc{PowerModelsDistribution} 0.11.4~\cite{PMD_PSCC}.

\section{Numerical Results - IEEE ELTF}
\subsection{Line Length Estimation Results}

Fig.~\ref{fig:lle_pf} shows the absolute difference between the bus voltages from PF calculations with the original and the estimated lengths. Very small differences are achieved rather quickly, and with 200 time steps the difference is almost never above 0.001\,p.u. (0.23\,V). This value is even lower than the standard deviation of the SM's random noise ($\approx$0.38\,V), so the calculated model appears suitable to be adopted in the practice. Due to the existence of multiple solutions, individual line lengths cannot be detected accurately in general, as shown in Fig.~\ref{fig:wrong_lengths} for the 200 time steps case, where every dot corresponds to one of the feeder's lines. Cumulative R and X values are thus a better indication of the estimation quality, which is also quite good, as shown in Fig.~\ref{fig:lle_xr}. At 200 time steps, the mean error in the R and X estimation is very close to zero, with few outliers. Note that the higher the loads, the more these outliers would cause inaccurate PF results. However, at the same time, the higher the loads/power flows, the higher the voltage drops, which would improve the LLE process. Thus, in this sense, the more urgently LLE is needed, the easier it is to perform.

 \begin{figure}[t]
\includegraphics[width=0.49\textwidth]{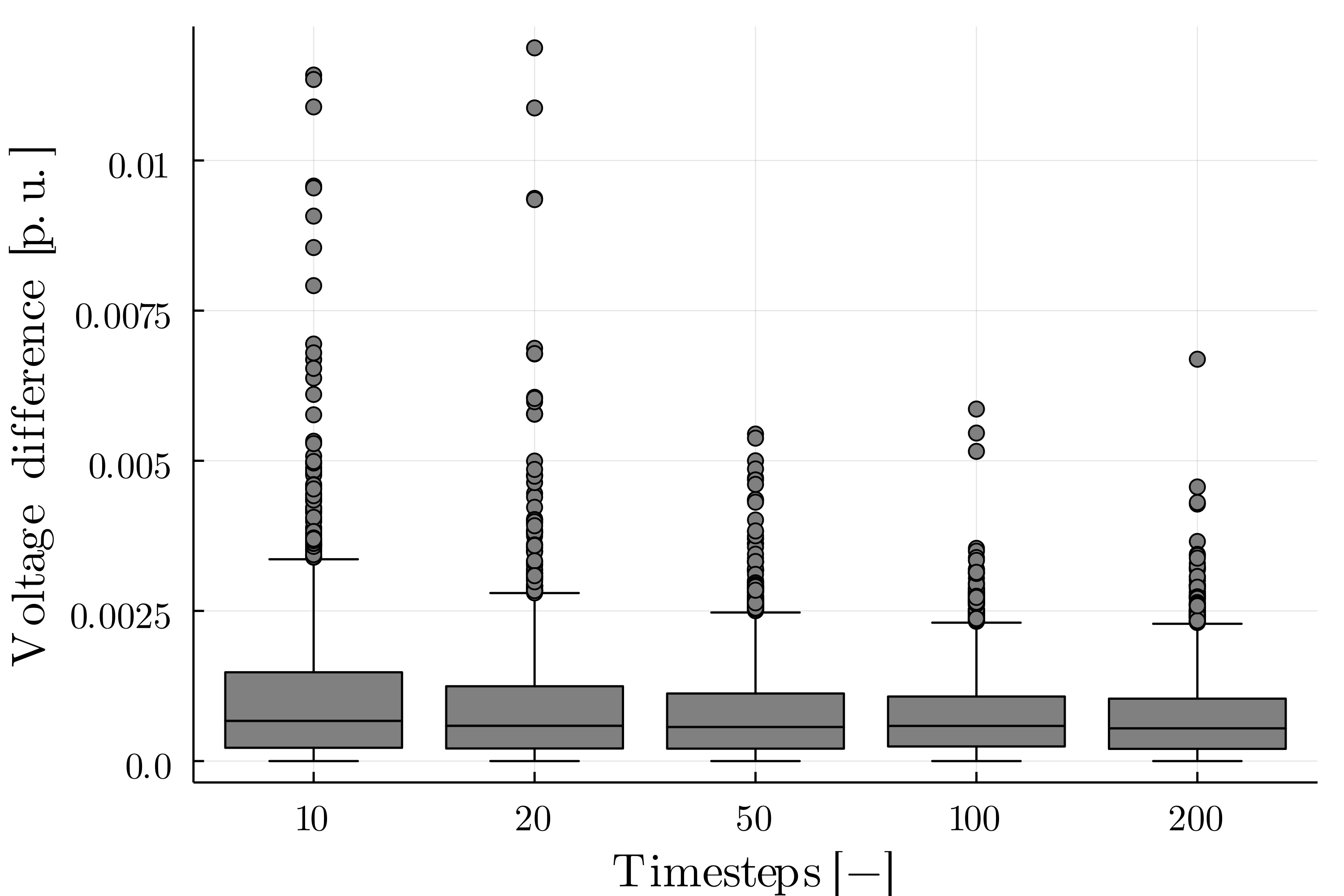}
\caption{Difference between the PF result with the original and estimated matrices (\textbf{transposed} IME case).}
\label{fig:ime_pf}
\end{figure}

\begin{figure*}[t!]
\begin{tabular}{cc}
  \includegraphics[width=0.45\textwidth]{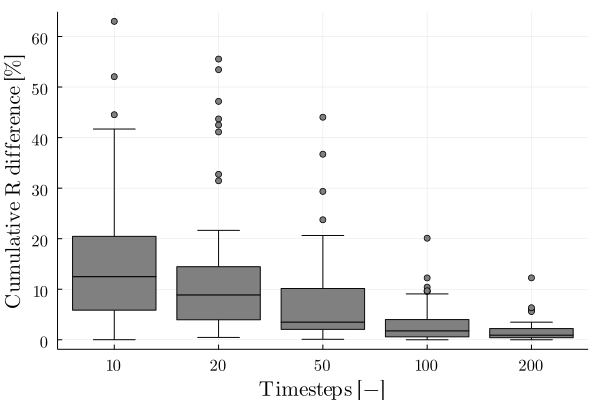}\label{fig:ime_R} &   \includegraphics[width=0.45\textwidth]{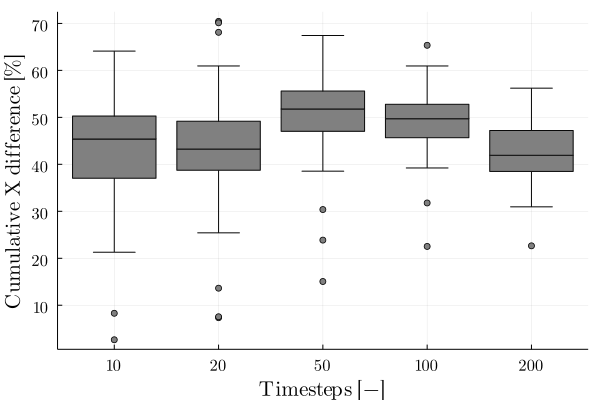}\label{fig:ime_X} \\
\end{tabular}
\caption{Difference between the users' actual and IME-calculated cumulative reactance (left) and resistance (right), \textbf{transposed} case. The boxplots have one entry for each user. Plot for $|$Z$|$ is omitted as it is similar to that of R (because X only contributes to a small part of $|$Z$|$).}
\label{fig:ime_rx}
\end{figure*}
 \begin{figure}[t]
 \centering
\includegraphics[width=0.45\textwidth]{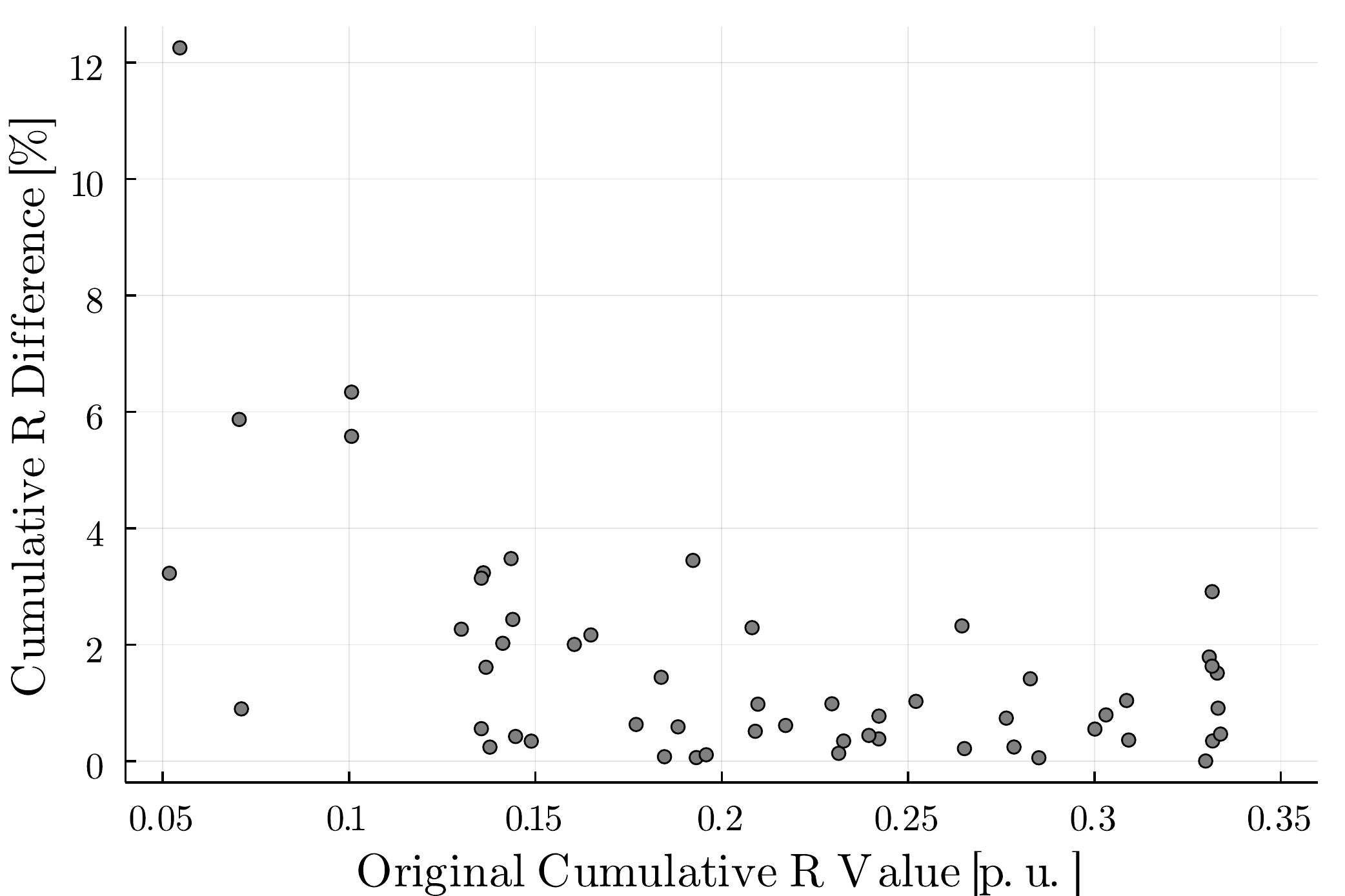}
\caption{\textbf{Transposed} IME result for 200 time steps, one dot per consumer. Poor accuracy outliers occur for low cumulative R values.}
\label{fig:r_per_user}
\end{figure}

\subsection{Impedance Matrix Estimation Results - Transposed Case}

Fig.~\ref{fig:ime_pf} is akin to Fig.~\ref{fig:lle_pf} but for the transposed IME case. The IME process is more computationally demanding and less accurate than LLE, which was to expect given the more pressing degeneracy. It can be observed that even with 200 time steps, while the third quartile of the boxplot in Fig.~\ref{fig:ime_pf} is below 0.001\,p.u., there are a few outliers of up to 0.007\,p.u. (1.61\,V). Nevertheless, the error seems small enough for the calculated model to be used in decision support applications (this is elaborated on in section~\ref{sec:discussion}).

Regarding the cumulative impedance values, Fig.~\ref{fig:ime_rx} shows that the identification of cumulative X values is more challenging than in the LLE case. In LLE, R and X are ``jointly" estimated, as they are both captured in the length, whereas here they are disjointed. As the X values are much lower than the R values and have a lower impact on the voltage drops, they are also more difficult to estimate. Fig.~\ref{fig:r_per_user} shows the error on the cumulative R estimation for each user at 200 time steps (these are the ``dots" in the boxplot in Fig.~\ref{fig:ime_rx}), showing a similar effect: the lower the absolute R values, the less accurate the IME, too. This is because, again, lower resistance values imply lower voltage drops (given the same power flow magnitudes). 

\subsection{Impedance Matrix Estimation Results - Untransposed Case}

\begin{figure*}[t]
\begin{tabular}{cc}
  \includegraphics[width=0.45\textwidth]{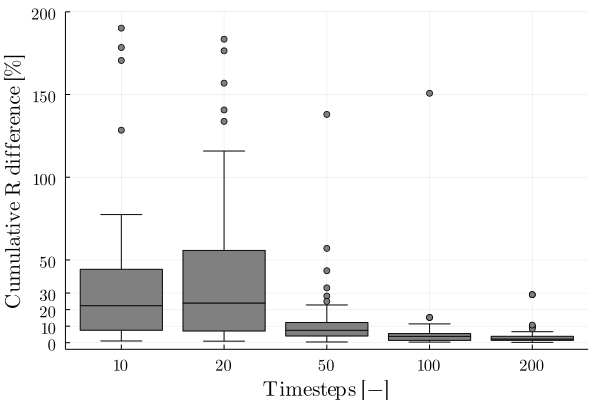}\label{fig:ime_R_untr} &   \includegraphics[width=0.45\textwidth]{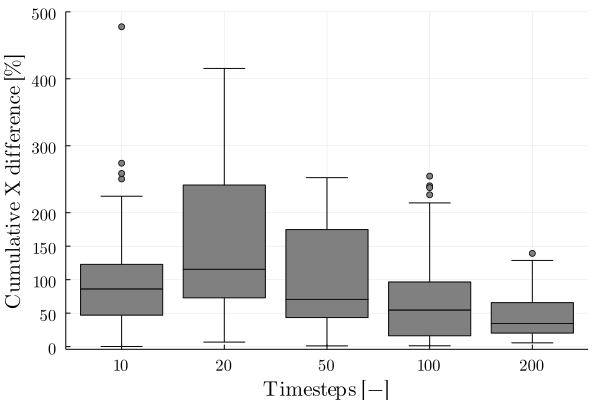}\label{fig:ime_X_untr} \\
\end{tabular}
\caption{Difference between the users' actual and IME-calculated cumulative reactance (left) and resistance (right), \textbf{untransposed} case. The boxplots have one entry for each user. Plot for Z is omitted as it is similar to that of R (because X only contributes to a small part of Z). Some outliers are omitted for ease of visualization. For R with 10 time steps: 828, 764, 506, 337, 282, 229. For R with 20 time steps: 466. For X with 10 time steps: 1250, 775, 774. For X with 20 time steps: 700.}
\label{fig:ime_rx_untr}
\end{figure*}

Fig.~\ref{fig:ime_pf_untr} shows the power flow validation for the untransposed IME case. An increased amount of outliers is observed for 10-20 time steps, with respect to the transposed case. Moreover, there is an anomalous bad performance for 20 time steps. This seems due to the measurement errors introduced in the additional 10 time steps added, which affect the R and X estimation as shown in Fig.~\ref{fig:ime_rx_untr}. In turn, the worse R and X estimations compromise the PF accuracy. Nevertheless, the results ultimately stabilize for 100-200 time steps. As a matter of fact, for 200 time steps, the performance is almost identical to that of the transposed case, both in terms of R estimation and power flow validation, as shown in Figs.~\ref{fig:ime_pf_untr},~\ref{fig:ime_rx_untr}. The estimation of cumulative reactance is worse than in the transposed case. Nevertheless, considering the modest X values of lines in  DNs, this does not significantly deteriorate the calculated impedance model. 

\subsection{Impedance Matrix Estimation Results - Diagonal Case}

\begin{figure*}[t]
\begin{tabular}{cc}
  \includegraphics[width=0.45\textwidth]{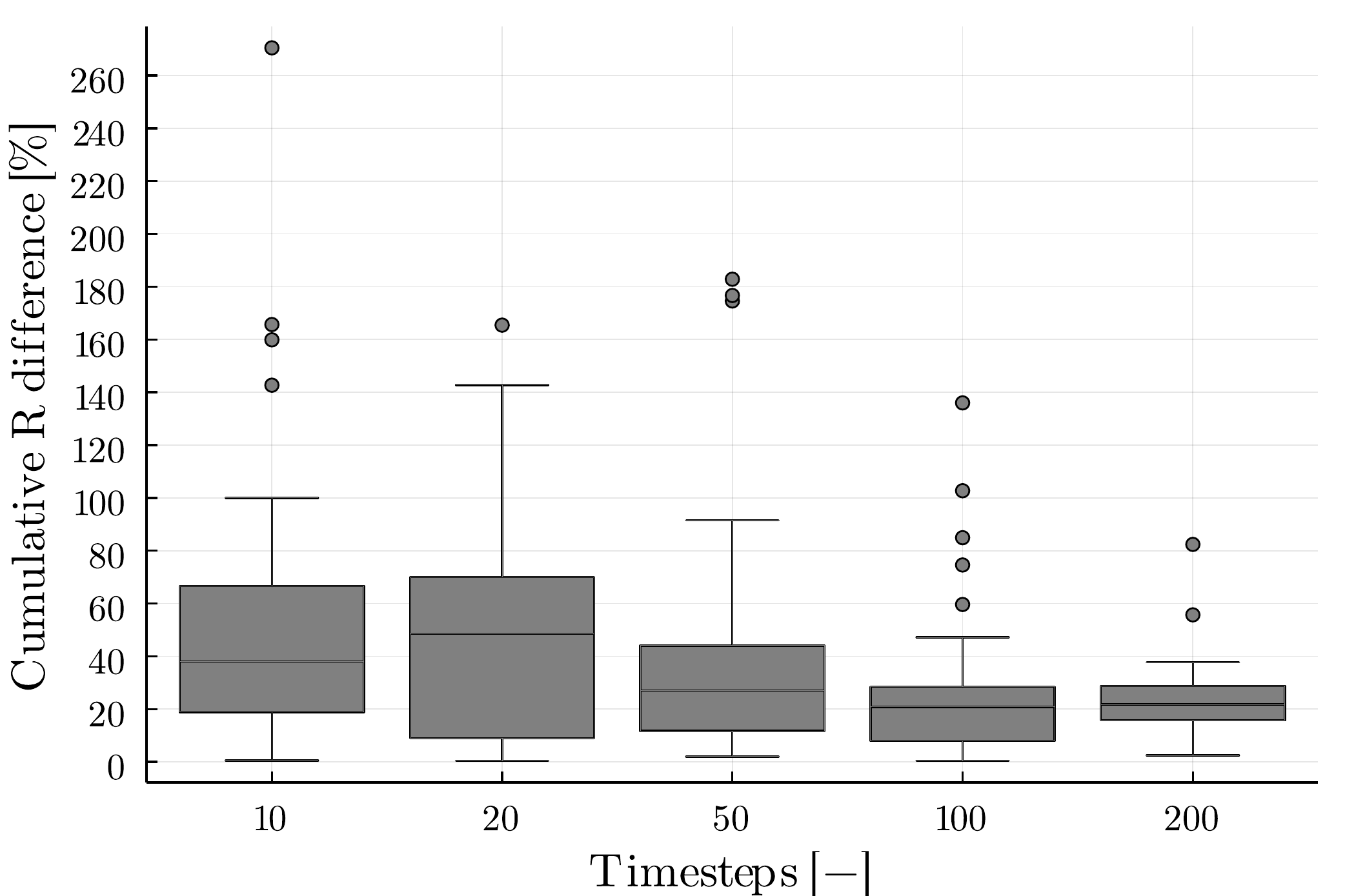}\label{fig:ime_R_diag} &   \includegraphics[width=0.45\textwidth]{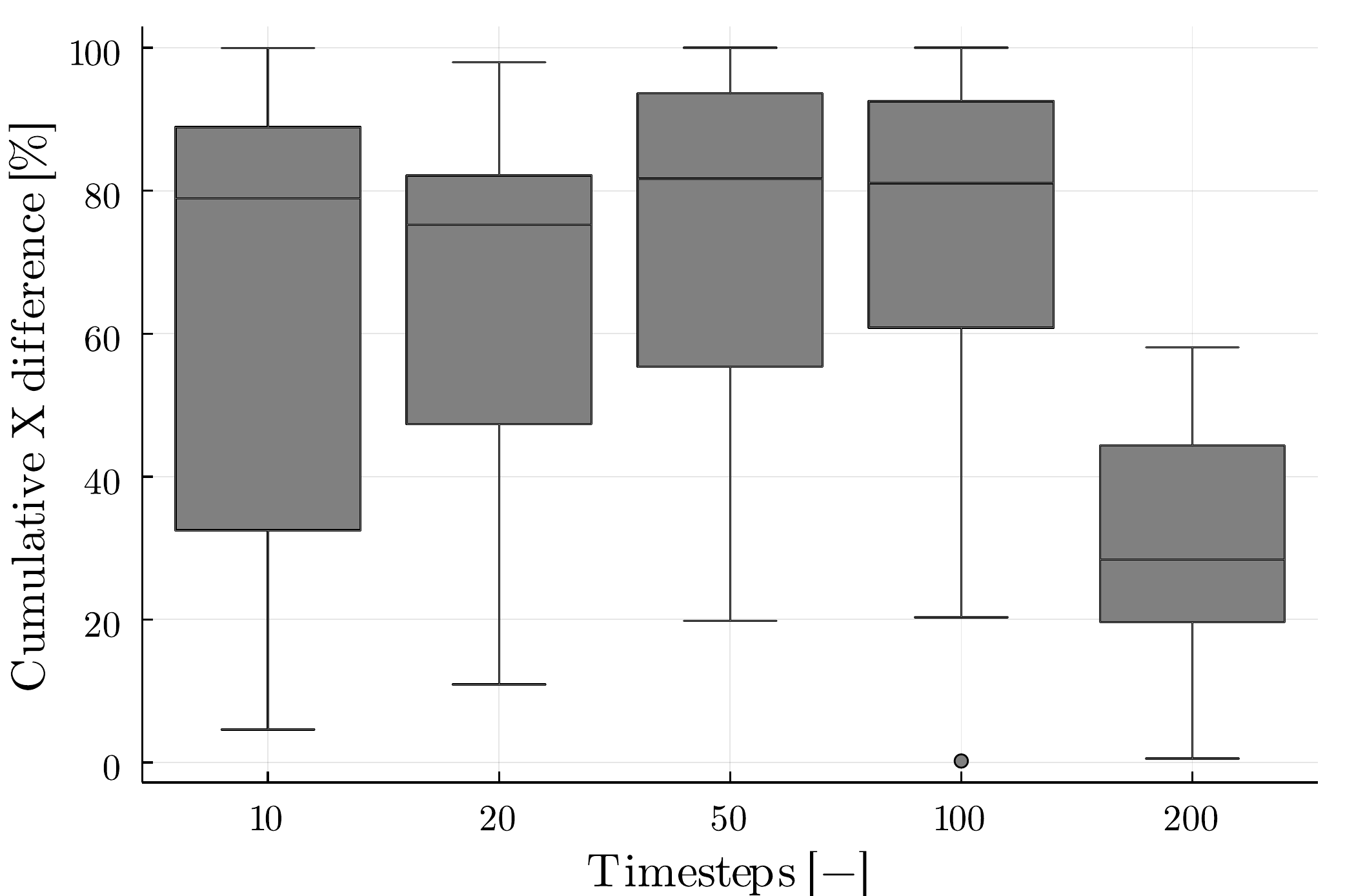}\label{fig:ime_X_diag} \\
\end{tabular}
\caption{Difference between the users' actual and IME-calculated cumulative reactance (left) and resistance (right), \textbf{diagonal} case. The boxplots have one entry for each user. Plot for Z is omitted as it is similar to that of R (because X only contributes to a small part of Z). }
\label{fig:ime_rx_diag}
\end{figure*}

\begin{figure}[t]
\includegraphics[width=0.49\textwidth]{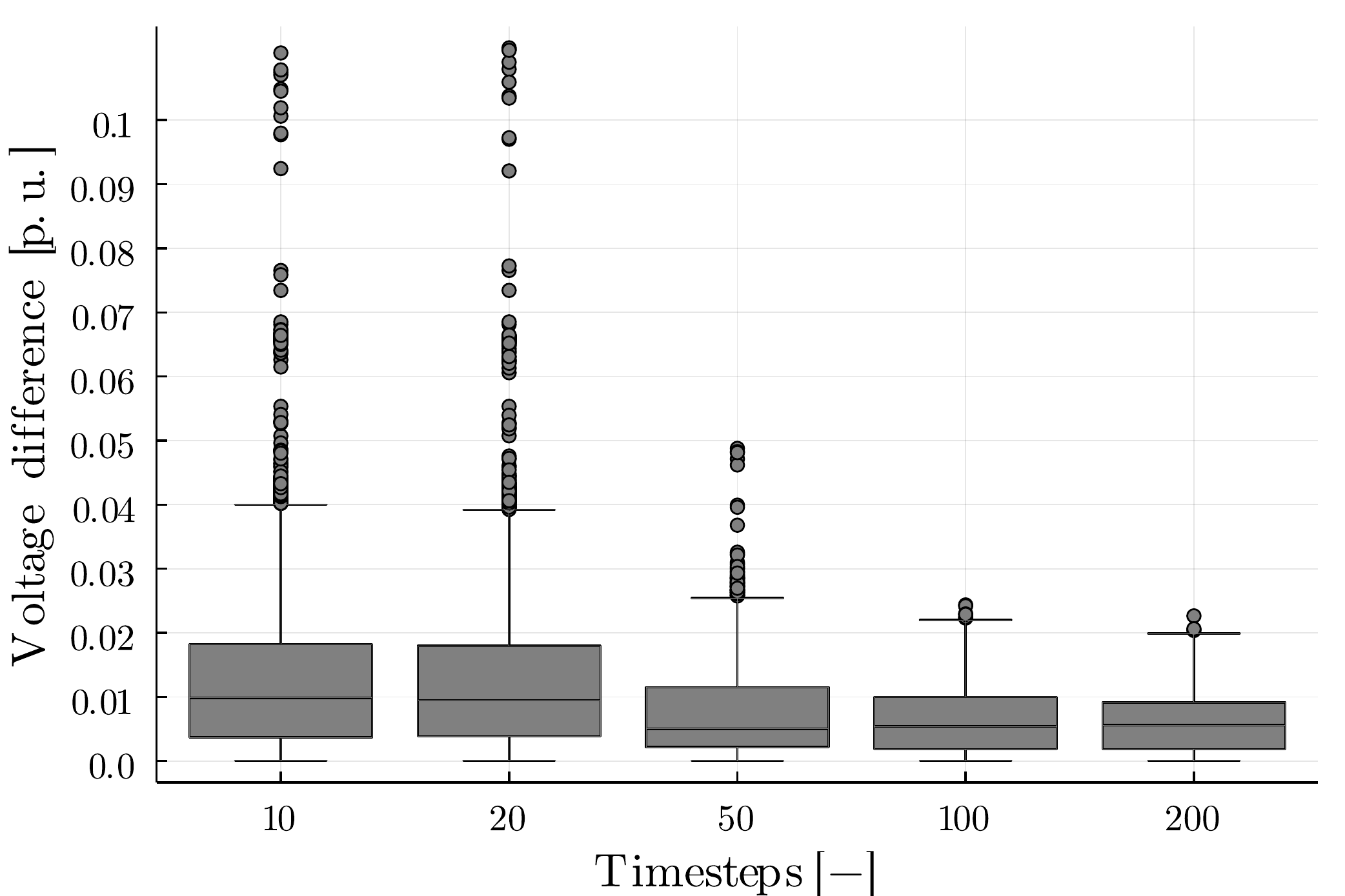}
\caption{Difference between the PF result with the original and estimated matrices (\textbf{diagonal} IME case).}
\label{fig:ime_pf_diag}
\end{figure}

Fig.~\ref{fig:ime_rx_diag}-\ref{fig:ime_pf_diag} show the R and X estimation and the power flow validation results for the diagonal IME case. For both analyses, the results are are worse than those of LLE and the other IME variants, with the exception of the cumulative X difference. However, the latter has a more modest impact on the overall impedance estimation and the power flow results. As it is the least realistic model, such performance is expected. A practical positive note, however, is that it is rather straightforward to infer realistic R/X ratios and upper and lower bounds for (self) R and X. In this case, the same R/X ratios as the previous cases have been used, whereas the R and X upper bounds have been tightened by a factor of 10: the upper bound of R for \textit{each line} now amounts to roughly 10$\times$ the largest total \textit{cumulative} R (and is as such still quite loose). This, together with the relatively limited number of variables and degeneracy, resulted in reduced computational time with respect to the other IME methods (see \ref{tab:comp_time}). Thus, there appears to be a substantial margin of improvement for the computational time of the other IME variants. However, calculating tighter bounds in presence of non-zero off-diagonal values appears less practical.

\subsection{Comparison and Discussion}\label{sec:discussion}

\begin{figure}[t]
\includegraphics[width=0.49\textwidth]{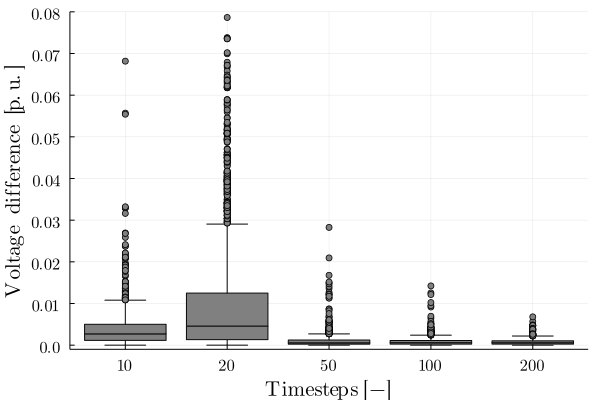}
\caption{Difference between the PF result with the original and estimated matrices (\textbf{untransposed} IME case). Two outliers at 10 time steps: 0.193, 0.113 are not displayed for ease of visualization.}
\label{fig:ime_pf_untr}
\end{figure}

The operating limits for LV DNs prescribed by standards like the EN\,50160 allow some fluctuation from the nominal voltage, i.e., $0.9\, \text{p.u.} \leq U^{\text{mag}}_i \leq 1.1 \, \text{p.u.} \, \,\, \forall i \in \busesm$. For decision support applications, e.g., OPF-based dispatch, one way to mitigate the impact of network model errors is to enforce tighter operating limits. However, this may result in higher costs or unnecessary renewable energy curtailment. The impedance models derived in all the presented cases have only few voltage outliers (see Figs.~\ref{fig:lle_pf}, \ref{fig:ime_pf}, \ref{fig:ime_pf_untr}), which are always less than 0.01 p.u. (except the diagonal IME case), i.e., well below the allowed $\pm$0.1 p.u. As such, the impedances derived with the proposed method are suitable for decision support purposes and can avoid the costs entailed by limit tightening.

Table~\ref{tab:comp_time} shows the computational time required by \textsc{Ipopt} to converge to a solution. 
In all cases, the number of power flow variables and constraints increases linearly with the number of time steps, implying increasing computational time. 
It can be observed that the LLE is, as expected, much faster than the (un)transposed IME, especially for cases with many time steps. 
The transposed IME is faster than the untransposed IME. This is also expected: computations are longer for the cases with higher amount of variables and possible solutions. The diagonal IME, in turn, outperforms both, but this is due to the ability tighten the variable bounds.

In any case, even the untransposed variant converges in a few hours, which is acceptable as these calculations are not expected to be repeated often. Note that, even though they are more computationally expensive, IME-derived models (particularly untransposed ones) are more generic, and might be the only viable estimation solution if cable type information is very rough or unavailable. Furthermore, they allow to derive improved 3$\times$3 impedance matrices for four-wire lines for which Kron or other reductions cause non-negligible approximation errors.

To retain the advantages of IME, i.e., not requiring cable/line type information, while managing the problems with the existence of multiple equivalent solutions, Carson's equations could be used to split the space of the impedance matrices into physically realistic and unrealistic ones. 
Carson's equations are nonlinear but continuous, so they are a candidate for inclusion in nonlinear optimization models.  

To achieve improved results and potentially remove the remaining outliers, more training samples could be used, ideally with more favourable characteristics, like increased voltage drops and/or differing voltage drop patterns. However, being able to estimate parameters with short time series is 1) an advantage of methods like the presented ones, that exploit non-approximated network physics to compensate for the limited measurement information, and 2) a desirable feature, as longer time series might be impractical to obtain with the present measurement infrastructure, especially in LV DNs. As an example, LV SM such as those used in Flanders have a memory buffer from which ten-day measurements can be queried remotely. Estimating the system parameters with this amount of data is, as such, desirable, as no additional measurement storage/handling strategies need to be deployed. As SM technologies in other regions are similar, and possibly from the same few manufacturers, similar considerations are expected to hold broadly.


\begin{table}[t]
\centering
\caption{Computational time to convergence to 1E-7, in seconds.}
\begin{tabular}{l  r  r  r  r  r} 
\hline
\# Time steps & 10 & 20 & 50 & 100 & 200 \\ 
\hline
LLE & 32 & 38 & 95 & 273 & 1\,774 \\ 

IME  transposed& 31 & 61 & 366 & 1\,271 & 10\,544 \\

 IME  untransposed& 52 & 174 & 7\,436 & 12\,106 & 13\,603 \\
 
 IME diagonal & 25 & 74 & 259 & 569 & 1\,454 \\
 \hline
\end{tabular}
\label{tab:comp_time}
\end{table}

\section{Numerical Results - Real LV Feeder}

The previous section applies the proposed methods to publicly available network and user power data. These have been chosen as they are known to the research community, making results reproducible, and because they are representative of European LV networks. However, public data sets do not always allow to capture all the complexity of working with real data. In a real system there are more unknowns, quantities are more uncertain, errors are less Gaussian, and so forth. In this section, the same methods as before are applied to a real feeder. The feeder is a semi-rural LV feeder, for which the Flemish DSO, Fluvius, collected two sets of ten-day measurement data in the context of the ADriaN project\footnote{https://www.energyville.be/en/research/adrian-active-distribution-networks}. The measurements were made available to the EnergyVille-affiliated authors of this paper, together with the DSO's existing digital model of the feeder. The measurement set-up is identical to that of the previous section: user-bus-only fifteen-minute averages of P, Q, U$^{\text{mag}}$ are available.

One of the ten-day sets is used for training and one for validation purposes. The data collection length is not random: the DSO can presently access a ten-day memory buffer remotely, which is assumed to be standard in the future in Flanders. 

The feeder is four-wire with a single grounding of the neutral at the transformer, radially operated and has only eight users, making for a quite different setting with respect to the IEEE ELTF. The small size results in no computational bottlenecks, but the IE process might be compromised by the short cables (i.e., probable limited impedances) and limited power flow magnitudes. Note that feeders of this size are quite common in Flanders, where there are often 4-5 feeders per transformer, with about forty users per transformer in total. Seven of the eight consumers are small residential users, whereas one is larger. Three users have three-phase connections, the other ones are single-phase (and so are their service cables). The three-phase users have rooftop PV, but the collected measurements refer to January 2022 and there is little injection at that time of year. No significant power quality issues are present. 

\begin{figure}[t]
    \centering
\includegraphics[width=0.85\linewidth]{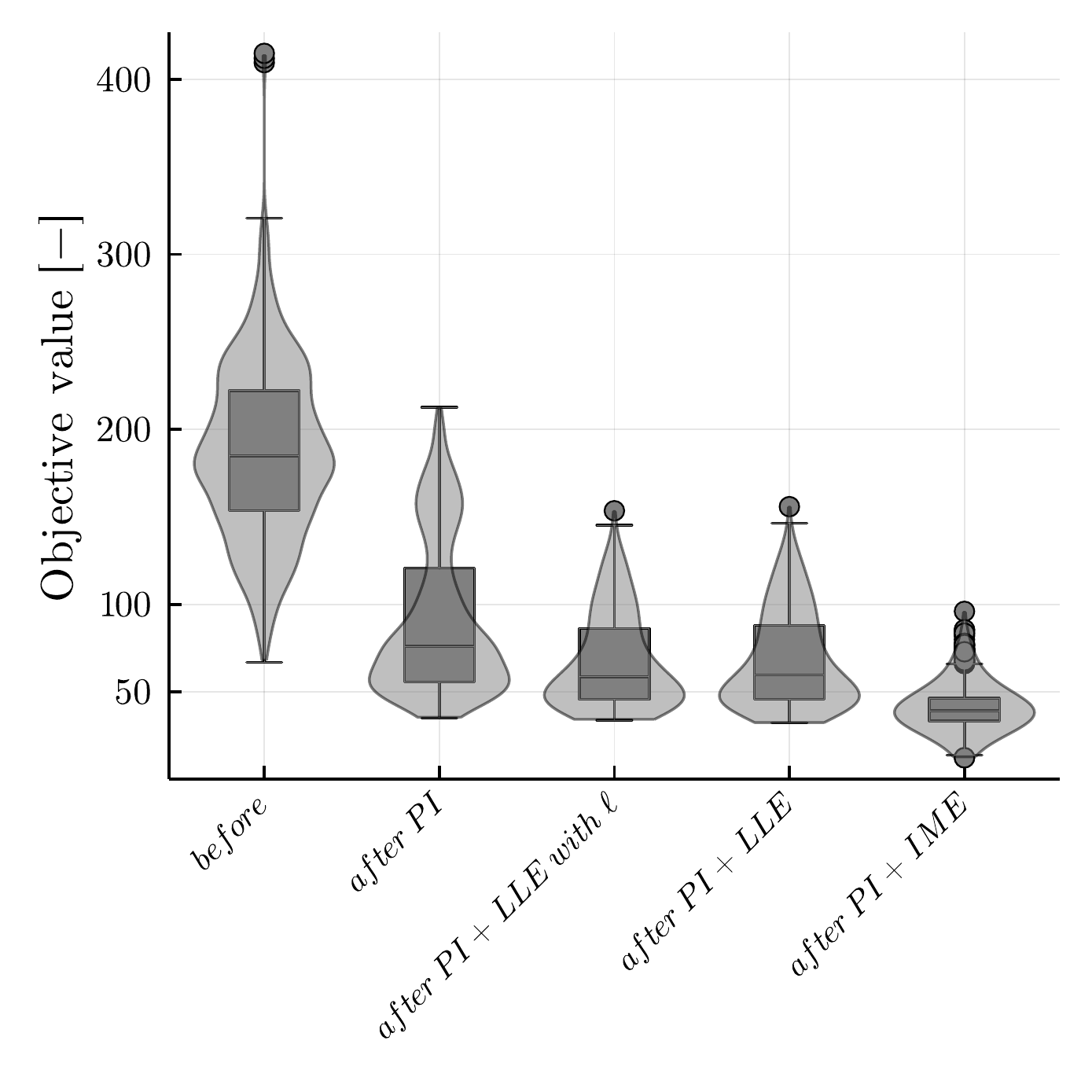}
    \caption{Objective values before and after network data corrections.}
    \label{fig:allin_fdr8}
\end{figure}

As the ``ground truth" network parameters are not known, it is not possible to validate the effectiveness of the proposed IE methods with certainty, as done before for the ELTF. This is hardly ever possible in real networks, where cables cannot be dug out, and more invasive measurement acquisition is not possible. Thus, validation techniques that are workable in real-life are also needed. One such technique is state estimation, as explained in~\cite{VaninMagazine}, and is the validation technique of choice here. Fig.~\ref{fig:allin_fdr8} shows the objective of DSSE calculations using the measurements from the validation set (box/violin plots have a ``dot" for every time step in the ten-day interval, excluding time steps with missing values). Note that the objective value of DSSE is the sum of all measurement residuals, and is tantamount to the result of \eqref{eq:objective}-\eqref{eq:g} if impedance values are fixed (not problem variables), and without all the extra LLE/IME constraints described in the same section. Each box/violin plot in Fig.~\ref{fig:allin_fdr8} refers to a different condition of the feeder model: \textit{before} is as originally given by the DSO, \textit{after PI} is after phase identification (PI) is performed and the phase connectivity of the original method is consequently adapted\footnote{PI is performed because the original model had wrong connectivity information. When done sequentially,  PI needs to be performed prior to IE. We reported some PI discussion for completeness, but the details of the latter are beyond the scope of this paper.}. Similarly, \textit{after PI+LLE with $\ell$} and \textit{after PI+LLE} display the SE results after the branch lengths are corrected, with and without the addition of $\ell$ residuals\footnote{In this case, we assign $\underline{\ell}$ as 1\% of the DSO's GIS-based length, and $\overline{\ell}$ as 200\% of the same (see eq. \eqref{eq:llikelihood1}-\eqref{eq:llikelihood2}).}, respectively, and \textit{after PI+IME} refers to using a DN model with the matrices derived from IME. In this test case we only apply the most generic and realistic untransposed IME. Obviously, for the LLE the per-km impedance values as present in the original feeder data are used. 

\begin{figure}[b]
    \centering
\includegraphics[width=0.9\linewidth]{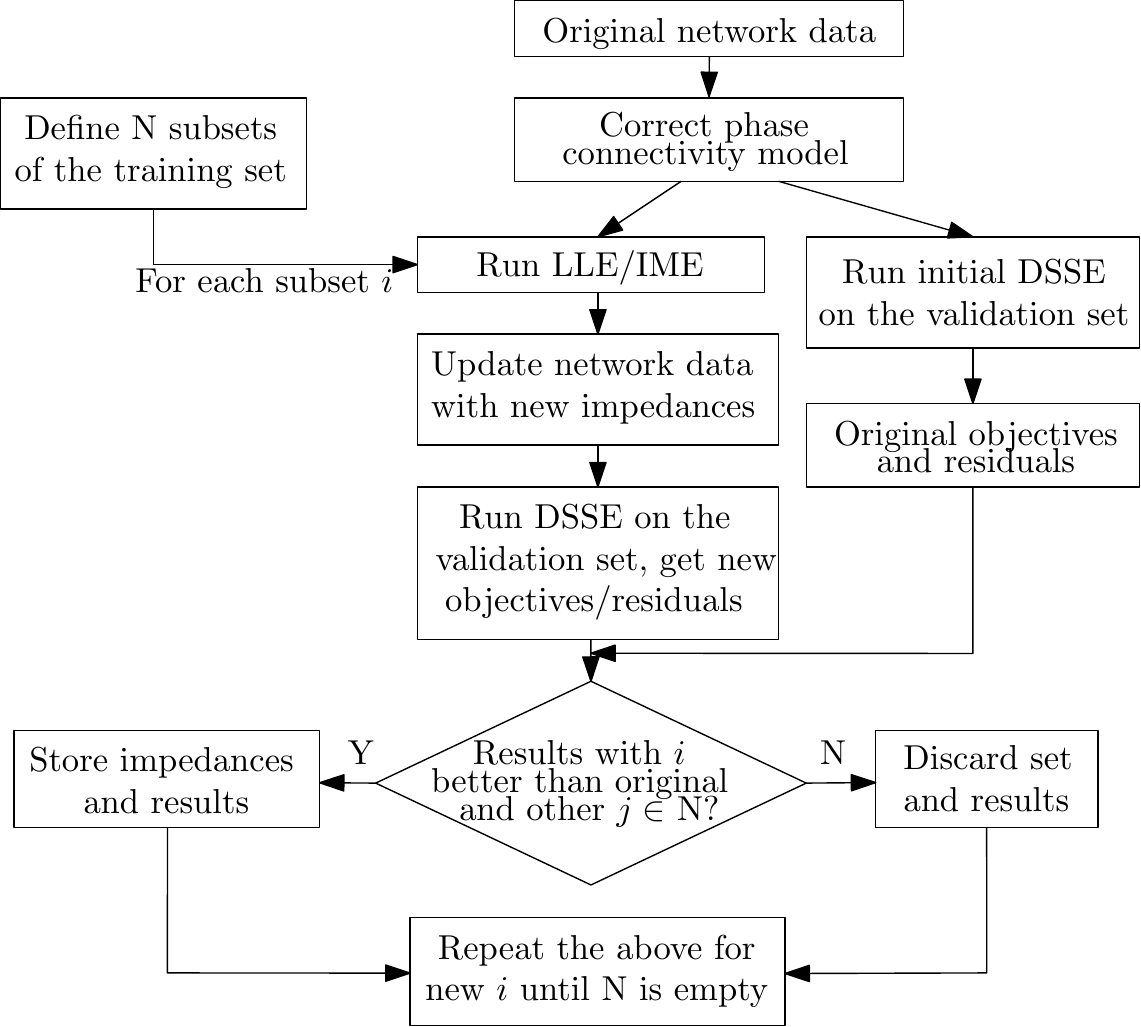}
    \caption{Illustration of the training set selection process.}
    \label{fig:campaign_flowchart}
\end{figure}

The lower the residuals, the higher the confidence that there are limited/no errors in the network models, which is the principle behind performing such validation with DSSE~\cite{VaninMagazine}. As such, Fig.~\ref{fig:allin_fdr8} shows promising results: SE residuals significantly reduce after DN model corrections, and the reduction is statistically consistent across all the validation set measurement time steps. Note that adding the $\ell$ to the residuals does not make a significant difference, and, as expected, IME performs better than LLE. On average, the objective values after PI+LLE are 23.2\% lower than  with PI only. With PI+IME, these values increases to 48\%. The \textit{individual} voltage residuals of the users (i.e., the difference between the voltage value measured at their SM, and the one estimated by the SE) are always well below 1\,V, and rarely exceed 0.5\,V, indicating high confidence in the calculated network model. 

Note that the training set used as input of the LLE and IME is only a subset of the given ten-day period, and the choice of the training set has been based on a trial-and-error process, shown in the flowchart in Fig.~\ref{fig:campaign_flowchart}. Firstly, a candidate set $i$ is tried, and the resulting IME/LLE outcome is used to modify the network data. Secondly, SE is run on the validation set, with the new network model. Finally, it is checked whether the new SE residuals/objective values effectively improved on the previous results. 

In practice, for this exercise, the number of subsets was thirteen: the ten time steps with higher loading form the start. The subsequently ten most loaded time steps are increasingly added, until 130 time steps are reached. The scenario with the 100 most highly loaded time steps has finally been picked as the one with the best results: adding extra time steps introduced more noise than added value. Understanding the relationship between the features of measurement data and the effectiveness of IE methods is an interesting field for future research. However, the simple physical intuition detailed above (i.e., larger loading equals more information) is already a good starting point. Note how this process allows to pick a favourable time window from the original measurement set. This limits the impact of bad data to some extent. Additional research is required to further identify the interactions between jointly present measurement and modelling uncertainties/unknowns~\cite{nyserda}, but is out of the scope this work. However, it is worth noting that the ability to solve IE with limited measurement data (little more than a day is required in the present exercise) may allow to discard a relatively large amounts of suspicious/corrupted measurements without worrying about being left with too few. The same is also useful when dealing with missing data: for this network, all the time steps that contained even a single missing measurement were  discarded, to avoid unfavourable measurement conditions. Finally, (W)LAV estimators like the ones used in the presented method are known to have excellent bad data rejection properties~\cite{bible}.

%
%
\section{Conclusions}\label{sec:conclusions}
This paper presents a novel method to combine state and line impedance estimation in unbalanced networks.
We note that the approach also allows one to run state estimation based purely on topology, with impedance being derived implicitly from the measurement history\footnote{Note that this does not hold for ``single-time-step" SE/IE, but only if done on a history of measurements. Solving SE with variable impedance for each time step independently is prone to overfitting, i.e., the solver may trivially minimize the WLAV residuals by fine-tuning the impedance values.}. 
The proposed approach solves a non-convex quadratic constrained optimization problem, relying on exact nonlinear steady-state power flow physics, and can take both the form of line length and impedance matrix estimation. 
For the latter case, constraints are added to impose structural properties to the matrices. 
In this paper, multi-conductor impedance matrices have size $3 \times 3$. This allows to appropriately model three-phase three-wire networks (typical MV networks), multi-grounded three-phase four-wire networks\footnote{Through Kron's reduction of the neutral.}, e.g., Australian LV networks, and single-grounded three-phase four-wire networks\footnote{Through the phase-to-neutral transformation.}~\cite{GethACM}, e.g., Belgian LV networks. 
Extending the method to either explicitly model three-phase four-wire lines instead of reduced models, possibly including Carson's equations, is left for future work.  


Despite the degeneracy and non-convexity, impedance identification is performed in acceptable computational time, and achieves good estimations with relatively short measurement time series: approximately two days of collected data. 
The length estimation is, as expected, faster and more accurate than the more general and degenerate matrix estimation, given the same number of time steps. 
Nevertheless, the latter also shows good results, and is a reliable alternative when length estimation cannot be used, e.g., if the line/cable types are unknown.
All the derived impedance models are accurate enough to be used for decision support applications, perhaps with the exception of the diagonal matrix model. 
To further improve the results, larger and ad-hoc selected training samples can be used. 
Finally, it is reassuring to note that both length and matrix estimations are more accurate when more needed: when impedance values or voltage drops are higher.

\bibliographystyle{IEEEtran}
\bibliography{IEEEabrv,Bibliography.bib}

\end{document}